\documentclass[useAMS,usenatbib]{mn2e}
\usepackage{amsmath}
\usepackage{graphicx}
\title[CR/X-ray effect on SMBH formation
]{Effect of cosmic ray/X-ray ionization on supermassive black hole formation
}
\author[K. Inayoshi and K. Omukai]{Kohei Inayoshi$^{1}$\thanks{E-mail:
inayoshi@tap.scphys.kyoto-u.ac.jp} and Kazuyuki
Omukai$^{1}$
\thanks{E-mail: omukai@tap.scphys.kyoto-u.ac.jp}
\\
$^{1}$Department of Physics, Kyoto University, Kyoto 606-8502, Japan }
\begin{document}

\date{}

\pagerange{000--000} \pubyear{0000}
%\pagerange{\pageref{firstpage}--\pageref{lastpage}} \pubyear{0000}

\maketitle

\label{firstpage}

\begin{abstract}
We study effects of external ionization by cosmic rays (CRs) and X-rays 
on the thermal evolution of primordial clouds under strong far-ultraviolet (FUV) 
radiation.
A strong FUV radiation field photodissociates H$_2$ and quenches its cooling.
Even in such an environment, a massive cloud with the virial temperature $\ga 10^{4}$ K
can contract isothermally at $8000$ K by the hydrogen Lyman $\alpha$ cooling.
This cloud collapses monolithically without fragmentation, 
and a supermassive star ($\ga 10^5~{\rm M}_{\sun}$) is believed 
to form at the center, which eventually evolves to a supermassive black hole (SMBH).
However, candidates of FUV sources, including star-forming galaxies, are probably 
sources of strong CRs and X-rays, as well.
We find that the external ionization promotes H$_2$ production and 
elevates the threshold FUV intensity $J_{\rm crit}$ needed 
for the SMBH formation for CR energy 
density $U_{\rm CR} \ga 10^{-14}~{\rm erg~cm^{-3}}$ or X-ray intensity 
$J_{X} \ga 10^{-24}~{\rm erg~s^{-1}~cm^{-2}~sr^{-1}~Hz^{-1}}$ at 1 keV. 
The critical FUV flux increases as $J_{\rm crit} \propto {U_{\rm CR}}^{1/2}$ 
($\propto {J_{\rm X}}^{1/2}$) in the high CR (X-ray, respectively) limit.
With the same value of FUV intensity at the Lyman limit ($13.6$ eV), 
the H$^{-}$ photodissociation rate, with threshold of $0.755$ eV, increases
and the H$_2$ abundance decreases with decreasing effective temperature of 
the FUV sources $T_{\ast}$. 
The lower value of $T_{\ast}$ thus results in the lower critical FUV flux 
$J_{\rm crit}$ at the Lyman limit.
Using an empirical relation between intensities of FUV and CRs/X-rays from nearby 
star-forming galaxies, we find that external ionization effect remarkably enhances 
the critical FUV flux for sources with $T_{\ast}$ as high as $10^{5}$ K
and composed of stars with $\la 100~{\rm M}_{\sun}$
to a level that is not realized in any halo.  
This indicates that, to induce SMBH formation, the FUV sources must be
either Pop II/I galaxies with low brightness temperature ($T_{\ast}\sim 10^4$ K), 
Pop III galaxies ($T_{\ast}\sim 10^5$ K) with a very top-heavy IMF, or 
Pop III galaxies too young to harbor sources of CRs/X-rays, for example, 
supernova remnants or high-mass X-ray binaries.
\end{abstract}


\section{Introduction}
The origin of supermassive black holes (SMBHs), ubiquitously existing 
at the centers of present-day galaxies, still remains a mystery. 
Among them, the existence of the luminous SDSS quasars with estimated 
SMBH masses of a few times $10^9~{\rm M}_{\sun}$ by their luminosities $\ga 10^{47}$ erg s$^{-1}$
at redshift $z \ga 6$, when the age of Universe is less than 1 Gyr, (e.g., Fan 2006) poses serious constraints on their formation and evolution scenarios. 
Several authors have studied models for the SMBH growth by merger and gas accretion (e.g., Haiman \& Loeb 2001; 
Volonteri, Haardt \& Madau 2003; Li et al. 2007), starting from stellar-mass black holes (BHs), 
which are remnants of first generation of stars. 
However, various negative feedbacks prohibit their rapid growth. 

In a merger event, the BH experiences strong recoil, 
with typical velocity $\ga 100~{\rm km~s^{-1}}$, 
owing to gravitational wave emission (Herrmann et al. 2007; Koppitz et al. 2007).
The recoil velocity depends on the spin configuration of the merging pair and   
can reach as high as $\sim 4000~{\rm km~s^{-1}}$ for the anti-parallel spins 
(Campanelli et al. 2007). 
In high-redshift universe, typical halos tend to be low-mass and 
their escape velocity ($\la 10~{\rm km~s^{-1}}$) is smaller than the recoil velocity.
If BHs are ejected from their host halos in merging events, 
the BH growth process must repeat again from scratch.
On the other hand, in the gas-accretion scenario, the accretion rate onto the BH 
is at most similar to the Eddington limit. 
In this case, it takes $\sim 1$ Gyr to form a SMBH with $10^9~{\rm M}_{\sun}$ 
from a BH with $100~{\rm M}_{\sun}$. 
Although this is marginally consistent with the existence of SMBHs at $z \ga 6$, 
this requires constant supply of gas at the Eddington rate.
However, recent hydrodynamical simulations demonstrate that 
radiative feedback from the the accreting black hole prevents such efficient 
accretion (Milosavljevi{\'c}, Couch \& Bromm 2009; Alvarez, Wise \& Abel 2009). 

We here explore another possibility that the seed BHs form 
via gravitational collapse of supermassive stars (SMSs) and thus 
are as massive as $\ga 10^{5}~{\rm M}_{\sun}$ from the beginning.  
Evolution and general relativistic instability of the supermassive stars have 
been investigated by many authors 
(e.g., Chandrasekhar 1964a, b; Zeldovich \& Novikov 1971; Shapiro \& Teukolsky 1983). 
Among them, Shibata \& Shapiro (2002) calculated the collapse of 
a rotating supermassive star into a SMBH and 
found that most of the mass originally in the SMS is eventually locked in the 
BH ($M_{\rm BH}\simeq 0.9~M_{\rm SMS}$). 

The biggest challenge in this scenario is how such supermassive stars 
are formed.
A promising pathway is through the collapse of a cloud in a halo under
an extremely strong far-ultraviolet (FUV) radiation field.  
The FUV radiation in the Lyman and Werner bands ($11.2-13.6$ eV) dissociates 
H$_2$ molecules, which are important coolants in a primordial gas 
in the low-temperature ($\la 8000$ K) regime.
Under a FUV radiation field exceeding a critical value,
$J_{\rm{crit}}\simeq 300~(10^5)$ in the unit of 
$10^{-21}~{\rm erg~s^{-1}~cm^{-2}~sr^{-1}~Hz^{-1}}$, 
for the stellar-type radiation with brightness temperature $T_*=10^4$ K ($10^5$ K, respectively), 
the H$_2$ formation is suppressed all the way during the protostellar 
collapse and then its cooling is totally quenched (Omukai 2001; Shang, Bryan \& Haiman 2010).
In this case, the collapse proceeds almost isothermally 
at $\simeq 8000$ K by hydrogen Lyman $\alpha$ and two-photon emission, and 
H$^{-}$ bound-free emission. 
Unlike the case with weaker radiation field, where the H$_2$ line cooling 
induces fragmentation to clumps of $\sim 10^3~{\rm M}_{\sun}$, 
such a cloud eludes fragmentation and collapses directly to a supermassive star of 
$\ga 10^{5}~{\rm M}_{\sun}$ (Bromm \& Loeb 2003; Regan \& Haehnelt 2009a, b), 
and collapses eventually into the seed BH for the subsequent growth to a SMBH.

The critical FUV flux for the suppression of the H$_2$ cooling is far
above the expected mean value of the background radiation.
Studying the spatial correlation between halos, 
Dijkstra et al. (2008) derived the probability distribution of the intensity 
of Lyman-Werner flux $J_{21}^{\rm LW}$ (in units of $10^{-21}~{\rm{erg}~
\rm{s}^{-1}~\rm{cm}^{-2}~\rm{sr}^{-1}~\rm{Hz}^{-1}}$) incident on halos 
with mass $\sim 10^8$ M$_{\sun}$ collapsing at redshift $z\simeq 10$. 
The estimated mean Lyman-Werner background is $J_{21}^{\rm LW}=40$, 
far below the critical value.
This means that such intense FUV fields are only realized in
neighborhoods of strong FUV sources.
They also estimated the fraction of halos bathed in radiation fields
exceeding the threshold value $J_{\rm crit}$.
This fraction is $10^{-6}$ for $J_{\rm crit}=10^3$ and decreases exponentially
with increasing $J_{\rm crit}$ ($\ga 2\times 10^3$).
A small difference in $J_{\rm crit}$ significantly affects the number of halos
bearing supermassive stars, and thus correct knowledge of the critical
FUV flux is crucial in estimating the number of seed BHs.

So far, the critical FUV flux $J_{\rm crit}$ has been studied only 
in cases where the incident radiation consists of components below 
the Lyman limit (Omukai 2001; Bromm \& Leob 2003; Shang et al. 2010). 
However, strong FUV sources, i.e., actively star-forming galaxies, are expected to 
contain a large number of massive stars and possibly some mini-quasars. 
Since massive stars end their lives as supernovae and 
leave the remnants, where cosmic rays (CRs) are accelerated and X-ray photons are 
produced, sources of strong FUV radiation can also be those of CRs and X-ray photons.
Similarly, mini-quasars and high-mass X-ray binaries emit soft X-ray radiation. 
The incident flux is thus expected to have such high-energy components. 
If present, CRs and X-rays enhance the ionization degree of gas and increase 
the amount of H$_2$, which is formed by electron-catalysed reactions.
In fact, Haiman, Rees, \& Loeb (1996) and Glover \& Brand (2003), who studied 
the condition for virialized minihalos under both FUV and X-ray irradiation
to be able to cool via H$_2$ line emission,  
found that the FUV photodissociation effect is somewhat alleviated
by the X-ray ionization.

In this paper, we consider similar effects of X-ray and CR ionization on  
more massive halos with the virial temperature $\ga 10^4$ K.  
Even without H$_2$ cooling, the clouds in such halos start collapsing via H atomic cooling.
If the FUV field is below some critical value $J_{\rm crit}$, however, 
abundant H$_2$ molecules form and fragmentation of the clouds occurs 
at some high density during the collapse by the H$_2$ cooling. 
With the enhanced H$_2$ fraction by X-ray and CR ionization, 
the more FUV flux is necessary to quench its cooling, thereby 
boosting the critical value $J_{\rm crit}$.
This results in fewer number of seed BHs  
since the isothermal collapse via atomic cooling is required to form SMSs. 
Our aim is to see the dependence of the critical FUV flux $J_{\rm crit}$ 
on the amount of CRs as well as X-rays.
By this way, we try to discuss the nature of the FUV sources needed to induce 
SMS formation. 
 
The organization of this paper is as follows. 
In Section 2, we explain our model of calculation, 
including chemical reactions and effects of FUV, CRs/X-rays. 
We present our results for the evolution of the clouds irradiated 
by FUV radiation and ionized by CRs/X-rays in Section 3, 
and give analytic interpretation for the obtained critical FUV flux in Section 4. 
In Section 5, we use an empirical relation between intensities of FUV and CRs/X-rays 
from nearby star-forming galaxies and speculate conditions 
to induce the supermassive star formation. 
Finally, we summarize this paper and give some discussions in Section 6. 

\section{Model}
\subsection{thermal evolution}
We consider thermal evolution of a metal-free cloud
in a moderately massive halo with virial temperature $\ga 10^{4}$ K
irradiated by a FUV field and simultaneously by either CRs or X-rays.
We suppose the radiation sources are extragalactic background 
or nearby other halos, rather than local ones in the same halo.  
In star-burst galaxies, although the internal radiation field 
can be very high, the interstellar media are expected to be metal-enriched.
Since even a low level of metal enrichment ($\ga 10^{-5}~Z_{\sun}$)
induces fragmentation and prevents the SMS formation (Omukai, Schneider 
\& Haiman 2008), we do not consider here the cases of internal sources. 

The evolution during the gravitational collapse is calculated
by a one-zone model (e.g., Omukai 2001).
The actual hydrodynamical collapse of self-gravitating clouds is well described by 
the Penston-Larson self-similar solution (Penston 1969; Larson 1969). 
In this solution, the clouds consist of two spatial parts: 
the flat central core, whose size is about the Jeans length $\lambda_{\rm J}$ and 
where number density of hydrogen nuclei $n_{\rm H}$, temperature $T$, 
chemical concentrations of species $i$, $y(i)$, etc. are nearly homogeneous, 
and the envelope, where the density profile obeys the power law with radius 
as $\rho \propto r^{-2}$.
The central density increases approximately at the free-fall rate. 
In our one-zone model, all the physical variables such 
as $n_{\rm H}$, $T$, etc. are intended to indicate those at the center of the core.
It has been confirmed that  
the thermal evolution of primordial clouds under strong FUV fields but without 
X-ray/CR ionization by the one-zone model reproduces very well 
that by the full three-dimensional calculation (Shang et al. 2010). 

Our main interest in this paper is put on the cases with extremely 
strong FUV radiation, where H$_2$ is totally dissociated 
and its cooling is quenched.
To start gravitational collapse, such halos must be as massive as with 
its virial temperature $\ga 10^4$ K, where the H atomic cooling is effective. 
This requires the total mass of the halo being $\ga 2 \times 10^7~{\rm M}_{\sun}$, 
corresponding to $\ga 4 \times 10^6~{\rm M}_{\sun}$ of gas for 
the virialization epoch $z_{\rm vir} \simeq 10$. 
Since the pressure effect is not important in such massive halos, 
the baryonic density in the clouds $\rho$ is assumed to increase 
with the free-fall timescale: 
\begin{equation}
\frac{d \rho}{dt}=\frac{\rho}{t_{\rm ff}}, 
\label{eq:dndt}
\end{equation}
where $t_{\rm ff}=\sqrt{3\pi /32G\rho }$.
In evaluating $t_{\rm ff}$, we neglect the contribution from the dark matter. 
The dark-matter gravity dominates that by the gas and shortens the collapse 
timescale only just after the virialization, where the temperature increases 
adiabatically in any case.
We confirmed from experiments that the omission of dark-matter gravity 
does not cause any noticeable deviations in temperature evolution 
as a function of density.

The temperature evolution is calculated by solving the energy equation: 
\begin{equation}
\frac{de}{dt}=-p\frac{d}{dt}\Big( \frac{1}{\rho}\Big) -\frac{\Lambda _{\rm{net}}}{\rho },
\end{equation}
where $e$ is the internal energy per unit mass, 
\begin{equation}
e=\frac{1}{\gamma_{\rm ad} -1}\frac{k_{\rm B} T}{\mu m_{\rm H}},
\end{equation}
$p$ is the pressure, 
\begin{equation}p=\frac{\rho k_{\rm B} T}{\mu m_{\rm H}},
\end{equation}
$\mu$ is the mean molecular weight, $k_{\rm B}$ is the Boltzmann constant, 
$\gamma_{\rm ad} $ is the ratio of specific heat 
and $\Lambda _{\rm net}$ is the net cooling rate per unit volume.
We set $\mu =1.2$ and $\gamma_{\rm ad}=5/3$ in our calculation. 
The net cooling rate consists of the following contributions:
\begin{equation}
\Lambda_{\rm net}=\Lambda_{\rm H}+\Lambda_{\rm H_2}+\Lambda_{\rm HD}
+\Lambda_{\rm chem}-\Gamma_{\rm CR}-\Gamma_{\rm X},
\end{equation}
where $\Lambda_{\rm H}$, $\Lambda_{\rm H_2}$ and $\Lambda_{\rm HD}$ are the 
radiative cooling rates by H atom, H$_2$ molecule 
and HD molecule, respectively, $\Lambda _{\rm chem}$ is the net cooling rate 
associated with chemical reactions, and $\Gamma _{\rm CR}$ and 
$\Gamma _{\rm X}$ 
are the heating rates by CRs and X-rays, respectively.
In the metal-free gas, H cooling by Ly $\alpha$ and two-photon emissions 
is effective at temperature $\ga 8000$ K, while 
H$_2$ molecule is an important coolant at lower temperature.
At even lower temperature $\la 150$ K, HD molecule becomes the dominant coolant.
Such low temperature environment is realized for example in the case of 
ionization by intense CRs or X-rays, which promotes H$_2$ formation 
via electron catalysed reactions, and thereby lowering the temperature below 
$\simeq 150$ K. 
The H atomic cooling rate is calculated by solving the level populations 
as in Omukai (2001), while, for the H$_2$ and HD cooling rate, 
we adopt the fitting formulae by Galli \& Palla (1998). 
In our calculation, we neglect the radiative cooling by He species, 
which is only effective for $T\ga 10^5$ K. 
The methods of calculation for heating rates by CRs and X-rays are described 
in Section 2.4.2 and 2.4.3, respectively.

When the density increases and the cloud becomes optically thick, 
the intensities of external radiation, i.e., FUV, CRs and X-rays, 
are attenuated before reaching the center and also 
the radiative cooling by e.g. H Ly $\alpha$ and H$_2$ becomes ineffective 
by the self-absorption. 
We assume the radius of the central core $R_{\rm c}$ is a half of the 
Jeans length $\lambda_{\rm J}$; 
\begin{equation}
R_{\rm c}=\frac{\lambda_{\rm J}}{2}=\frac{1}{2}
\sqrt{\frac{\pi k_{B}T}{G\rho \mu m_{\rm H}}}.
\label{eq:rc}
\end{equation}
Since we focus on the center of collapsing clouds, the column number density of 
$i$-th chemical species is 
\begin{equation}
N(i)=y(i)n_{\rm H}R_{\rm c}.
\label{eq:opt}
\end{equation}  
We use this column density in considering the optical-depth effect 
on the incident radiation intensities, as well as on the radiative cooling rates.  
 
\subsection{chemistry}
We consider the chemical reactions in primordial gas among 
the following 14 species: 
H, H$_2$, e$^{-}$, H$^+$, H$_2^+$, H$^-$, D, HD, D$^+$, HD$^+$, D$^-$, He, He$^+$, and He$^{++}$.
The included reactions are summarized in Table~1.
The following points are worth noting.
We adopt the H$_2$ collisional dissociation rate (reaction 7) of 
Martin, Schwarz \& Mandy (1996), which is ten times larger than the older rate of
Shapiro \& Kang (1987) used by Omukai (2001)
around $\simeq 10^3$ cm$^{-3}$. 
Shang et al. (2010) found that this difference affects the value of the critical 
FUV flux $J_{\rm{crit}}$ by about an order of magnitude.
For the photoionization of H (reactions $21-25$)
and the photodissociation of H$_2^+$ (reaction 20),  
those from the excited levels are also included as in Omukai (2001).
\begin{table}
\begin{center}
\caption{Chemical Reactions}
\begin{tabular}{l c r}
\hline
Number & Reaction & Reference \\ \hline \hline
 & H collisional reactions & \\
\hline
1  &   H        +     e$^-$ $\rightarrow $ H$^+$ +      2e$^-$   &  1\\
2  &   H$^+$  +     e$^-$ $\rightarrow $ H       +    $\gamma $  &  3$^{\ast}$\\
3  &   H        +     e$^-$ $\rightarrow $ H$^-$ +    $\gamma $&  2\\
4  &   H$^-$  +    H       $\rightarrow $  H$_2$ +          e       &  2\\
5  &   H        +    H$^+$ $\rightarrow $  H$_2^+$ + $\gamma $&  2\\
6  &   H$_2^+$+   H        $\rightarrow $ H$_2$ +       H$^+$    &  2\\
7  &   H$_2$  +    H        $\rightarrow $ 3H                          &  4\\
8  &   H$_2$  +     H$^+$ $\rightarrow $ H$_2^+$ +       H       &  2\\
9  &   H$_2$  +     e$^-$ $\rightarrow $ 2H      +     e$^-$      &  2\\
10 &  H$^-$  +     e$^-$ $\rightarrow $ H        +     2e$^-$    & 1\\
11 &  H$^-$  +     H$^+$ $\rightarrow $ 2H                          &  2\\
12 &  H$^-$  +     H$^+$ $\rightarrow $ H$_2^+$ +   e$^-$     &  2\\
13 &  H$_2^+$ +   e$^-$ $\rightarrow $ 2H                          &  2\\
14 &  H$_2^+$ + H$^-$  $\rightarrow $  H$_2$ +      H           &  1\\
15 &  3H                     $\rightarrow $ H$_2$   +     H           &  5\\
16 &  2H       +   H$_2$ $\rightarrow $  2H$_2$                    &  5\\
17 &  2H$_2$              $\rightarrow $  2H       +     H$_2$     &  5\\
\hline
 & photo-, and CR reactions & \\ 
\hline
18 &  H$_2$  + $\gamma $ $\rightarrow $  2H                     &  6\\
19 &  H$^-$  + $\gamma $ $\rightarrow $ H    +      e$^-$    &  7\\
20 &  H$_2^+$   + $\gamma $ $\rightarrow $  H   +   H$^+$   &  8\\
$21-25$ &  H(n)    + $\gamma $ $\rightarrow $  e$^-$ +   H$^+$ ($n=1-5$)   &  9\\
26 &  H        +    CR     $\rightarrow $  e$^-$  +     H$^+$    &  10\\
\hline
 & D reactions & \\
\hline 
27 &  D$^+$ +   e$^-$ $\rightarrow $ D  +  $\gamma $         &  11\\
28 &  D       +   H$^+$ $\rightarrow $  D$^+$  +      H           &  12\\
29 &  D$^+$  +   H      $\rightarrow $  H$^+$   +     D           &  12\\
30 &  D       +   H       $\rightarrow $  HD    +   $\gamma $    &  11\\
31 &  D       +  H$_2$  $\rightarrow $  H       +     HD            &  11\\
32 &  HD$^+$  + H     $\rightarrow $  H$^+$  +     HD           &  11\\
33 &  D$^+$  + H$_2$ $\rightarrow $  H$^+$  +     HD           &  13\\
34 &  HD     +   H      $\rightarrow $  H$_2$  +      D            &  11\\
35 &  HD    +  H$^+$  $\rightarrow $  H$_2$  +      D$^+$      &  13\\
36 &  D      +   H$^+$ $\rightarrow $  HD$^+$  +  $\gamma $ &  11\\
37 &  D$^+$ +   H      $\rightarrow $  HD$^+$  +  $\gamma $ &  11\\
38 &  HD$^+$ + e$^-$  $\rightarrow $  H       +     D           &  11\\
39 &  D      +   e$^-$  $\rightarrow $   D$^-$  +  $\gamma $ &  11\\
40 &  D$^+$  +  D$^-$ $\rightarrow $  2D                          &  11\\
41 &  H$^+$  +  D$^-$ $\rightarrow $  D     +       H            &  11\\
42 &  H$^-$ +   D      $\rightarrow $   H       +     D$^-$       &  11\\
43 &  D$^-$  +  H      $\rightarrow $   D       +     H$^-$       &  11\\
44 &  D$^-$  +  H      $\rightarrow $   HD       +    e$^-$       &  11\\
45 &  HD      + $\gamma \rightarrow $ H        +      D            &  6 \\
\hline
 & He reactions & \\
\hline 
46 &  He  +  e$^-$      $\rightarrow $   He$^+$       +    2e$^-$       &  1\\
47 &  He$^+$  +  e$^-$      $\rightarrow $   He       +    $\gamma $       &  1\\
48 &  He$^+$  +  e$^-$      $\rightarrow $   He$^{++}$       +    2e$^-$       &  1\\
49 &  He$^{++}$  +  e$^-$      $\rightarrow $   He$^+$       +    H$^+$  +$\gamma $     &  1\\
50 &  He$^+$  +  H      $\rightarrow $   He       +    H$^+$  + $\gamma $     &  14\\
51 &  He  +  H$^+$      $\rightarrow $   He$^+$       +    H       &  14\\
52 &  He  +  $\gamma $      $\rightarrow $   He$^+$       +    e$^-$       &  1\\
\hline
\end{tabular}
\end{center}
(1) Abel et al. (1997); (2) Galli \& Palla (1998); (3) Ferland et al. (1992; Case B); 
(4) Martin et al. (1996); (5) Palla, Salpeter \& Stahler (1983); (6) Wolcott-Green \& Haiman (2011); 
(7) John (1988); (8) Stancil (1994); (9) Rybicki \& Lightman (1979); 
(10) Stacy \& Bromm (2007); (11) Nakamura \& Umemura (2002); (12) Savin (2002); 
(13) Galli \& Palla (2002); (14) Yoshida et al. (2006).
\end{table}

\subsection{Initial condition}
The calculation starts at $n_{\rm{H}}=0.1$ cm$^{-3}$, 
with initial temperature $T=160$ K, appropriate for halos virializing at $z_{\rm vir}=10$, 
whose turnaround epoch corresponds to $z_{\rm turn}\simeq 17$ 
(Omukai et al. 2008).
The early stages at densities $n_{\rm H}=0.1$ cm$^{-3}$ and $1.0$ cm$^{-3}$
correspond to redshifts of $z\simeq 15.5$ and $11.5$, respectively.
The initial concentrations of electron, $y({\rm e^{-}})=2\times 10^{-4}$, and H$_2$ molecule, 
$y({\rm{H}}_2)=2\times 10^{-6}$, are taken from the values 
at $\sim 0.1$ cm$^{-3}$ of Bromm \& Loeb (2003).
We set the initial He concentration at $y({\rm He})=0.08$,
which corresponds to the He mass fraction, $Y_{\rm He}=0.24$.

\subsection{Incident FUV radiation, CRs, and X-rays}
\subsubsection{FUV radiation}
The incident FUV radiation field $J_{\rm FUV}(\nu)$ is assumed 
to have a diluted thermal spectrum, i.e.,   $J_{\rm FUV}(\nu) 
\propto B_{\nu}(T_{\ast})$,  
with brightness temperature $T_{\ast}=10^4$ or $10^5$ K, corresponding to that from 
assembly of metal-enriched stars or massive Pop III stars, respectively. 
The normalization of the intensity is set
at the Lyman-limit frequency $\nu_{\rm L}=13.6$ eV by 
$J_{21}=J_{\rm FUV}(\nu )/10^{-21}~{\rm erg~s^{-1}~cm^{-2}~sr^{-1}~Hz^{-1}}$.
For the same value of $J_{21}$ at the Lyman limit, 
the rate of H$_2$ photodissociation, which proceeds via absorption of 
photons of $11.2-13.6$ eV does not vary so much 
for different values of $T_{\ast}$. 
On the other hand, H$^-$ photodissociation rate, 
whose threshold energy (0.755 eV) is far 
below the Lyman limit, is significantly affected with change of $T_{\ast}$ 
even with the same $J_{21}$.
For example, this rate is $2\times10^{4}$ times higher 
for $T_{\ast}=10^4$ K than that for $T_{\ast}=10^5$ K with the same $J_{21}$.
Since H$^-$ is the intermediary in the H$_2$-forming reaction 
(reactions 3 and 4), 
the H$_2$ concentration depends sensitively on $T_{\ast}$ (see Section 4). 

The H$_2$ and HD are self-shielded against photodissociation for 
their column densities $\ga 10^{13}$ cm$^{-2}$. 
The HD shielding is also contributed by H and H$_2$.   
We use the shielding factors by Wolcott-Green \& Haiman (2011).

\begin{figure}
\begin{center}
\centerline{
\begin{tabular}{l l}
\rotatebox{0}{\includegraphics[height=59mm,width=80mm]{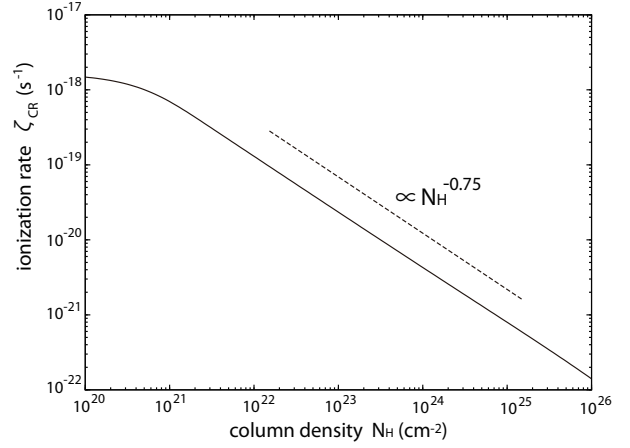}}
\end{tabular}
}
\caption{The cosmic-ray ionization rate $\zeta _{\rm CR}$ for $U_{\rm CR}=10^{-15}$ erg cm$^{-3}$ 
and $\epsilon _{\rm min}=10^6$ eV, and $\epsilon _{\rm max}=10^{15}$ eV
as a function of H column density $N_{\rm H}$. 
In the high column-density limit, the rate is approximated as $\zeta _{\rm CR}\propto N_{\rm H}^{-0.75}$.
}
\label{fig:zeta}
\end{center}
\end{figure}

\subsubsection{Cosmic Rays}
For the incident CR flux, we follow the treatment by Stacy \& Bromm (2007).
The CR energy distribution $dn_{\rm CR}/d\epsilon ~({\rm cm^{-3}~eV^{-1}})$ 
is assumed to obey a power-law spectrum with index $-2$, 
the value by the diffusive shock acceleration by a strong shock 
(e.g. Bell 1978) 
from the minimum energy $\epsilon _{\rm min}=10^{6}$ eV to the maximum 
$\epsilon _{\rm max}=10^{15}$ eV.
After incident to the gas, 
CRs propagate inward with losing their energies by ionization 
up to the penetration depth $D_{\rm p}$ at a rate $(d\epsilon /dt)_{\rm ion}$, 
for which we use the expression by Schlickeiser (2002).
The heating rate by CRs at depth $D$ is
\begin{equation}
\Gamma _{\rm{CR}}(D)=\frac{E_{\rm{heat}}}{\Delta \epsilon }\int ^{\epsilon _{\rm{max}}}_{\epsilon _{\rm{min}}}\Big( \frac{d\epsilon }{dt}\Big) _{\rm{ion}}\frac{dn_{\rm{CR}}}{d\epsilon }e^{-D/{D_p}}d\epsilon ,
\end{equation}
where $\Delta \epsilon =50$ eV is the approximate kinetic energy of a CR particle 
lost upon each scattering (Spitzer \& Tomasko 1968), 
and the ionization rate is 
\begin{equation}
\zeta _{\rm{CR}}(D)=\frac{\Gamma _{\rm{CR}}(D)}{n_{\rm{H}}E_{\rm{heat}}},
\end{equation}
where $E_{\rm{heat}}\simeq 6$ eV is the energy deposited as heat per ionization 
in a neutral medium (Spitzer \& Scott 1969; Shull \& van Steenberg 1985). 
Note that the CR ionization rate is 
related with the total CR energy density, 
\begin{equation}
U_{\rm{CR}}=\int ^{\epsilon _{\rm{max}}}_{\epsilon _{\rm{min}}}
\epsilon \frac{dn_{\rm{CR}}}{d\epsilon } d\epsilon ,
\end{equation}
by
\begin{equation}
\zeta _{\rm{CR}}=1.7\times 10^{-18}U_{15}\hspace{2mm}{\rm{s}}^{-1}, 
\end{equation}
in the low column density limit, where $U_{15}=U_{\rm{CR}}/10^{-15}$ erg cm$^{-3}$. 
In Fig.~\ref{fig:zeta}, we also plot the CR ionization rate $\zeta _{\rm CR}$ as a function of the column density $N_{\rm H}$. 
This can be approximated as
\begin{align}
\zeta _{\rm CR}&=1.3\times 10^{-19}U_{15}\Big( \frac{N_{\rm H}}{10^{22}~{\rm cm}^{-2}}\Big) ^{-3/4}~{\rm s}^{-1}, \nonumber\\
&=6.8\times 10^{-20}U_{15}\Big( \frac{n_{\rm H}}{10^3~{\rm cm}^{-3}}\cdot \frac{T}{8000~{\rm K}}\Big) ^{-3/8}~{\rm s}^{-1},
\label{eq:zeta}
\end{align}
for column density higher than $\ga 10^{22}~{\rm cm}^{-2}$.

The CR ionization rate in the Milky Way has been estimated by many authors, 
including $\zeta _{\rm{CR}}\simeq 4\times 10^{-16}$ s$^{-1}$ 
(Hayakawa, Nishimura \& Takayanagi 1961), $6.8\times 10^{-18}~{\rm{s}}^{-1}\la \zeta _{\rm{CR}} \la 1.2\times 10^{-15}~{\rm{s}}^{-1}$ 
(Spitzer \& Tomasko 1968) and 
$\zeta _{\rm{CR}}\simeq 3\times 10^{-17}$ s$^{-1}$ (Webber 1998). 
By recent observation of H$_3^+$ in the interstellar medium, 
the average H$_2$ ionization by CRs is evaluated  
$\zeta _{\rm{CR,H}_2}\simeq 4\times 10^{-16}$ s$^{-1}$ 
(McCall et al. 2003; Indriolo et al. 2007), which is translated 
to a rather high H ionization rate of 
$\zeta _{\rm CR} \simeq 2.6 \times 10^{-16}~{\rm s}^{-1}$. 
The CR intensity at high redshift is totally uncertain.
The CR energy density is set by a balance between the injection 
and the diffusive leakage.    
On one hand, star formation in young galaxies can be more active 
than in the Milky Way. 
The CR injection rate being proportional to the star formation rate, 
the gas in the neighborhood of such galaxies is subject to intense CRs.
On the other hand, magnetic fields are expected to be much weaker 
in young galaxies, 
which results in the longer Larmor radii of CRs and easier leakage from the 
galaxies.
This may result in weaker CR density if the CR sources are in the same halo. 
Considering those uncertainties, 
we here regard the CR energy density as a free parameter and 
calculate the cases for  
$10^{-3} \leq U_{15} \leq 10^4$, which corresponds to the CR ionization rate of 
$10^{-21}~{\rm s^{-1}} \la \zeta _{\rm CR} \la 10^{-14}~{\rm s^{-1}}$ in the low density case. 
\begin{figure*}
\begin{center}
\centerline{
\begin{tabular}{l l}
\rotatebox{0}{\includegraphics[height=59mm,width=80mm]{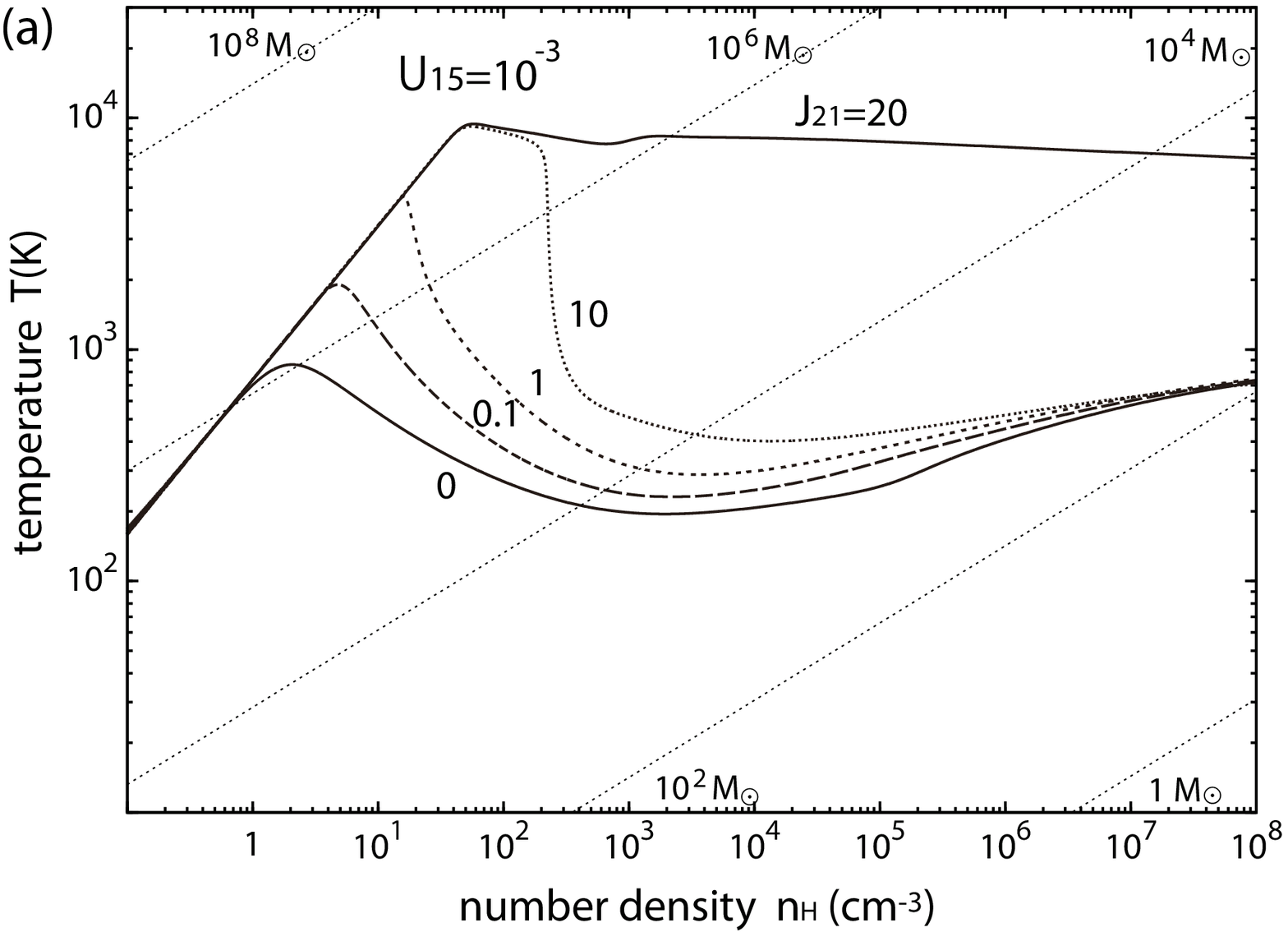}} \hspace{6mm}
\rotatebox{0}{\includegraphics[height=59mm,width=80mm]{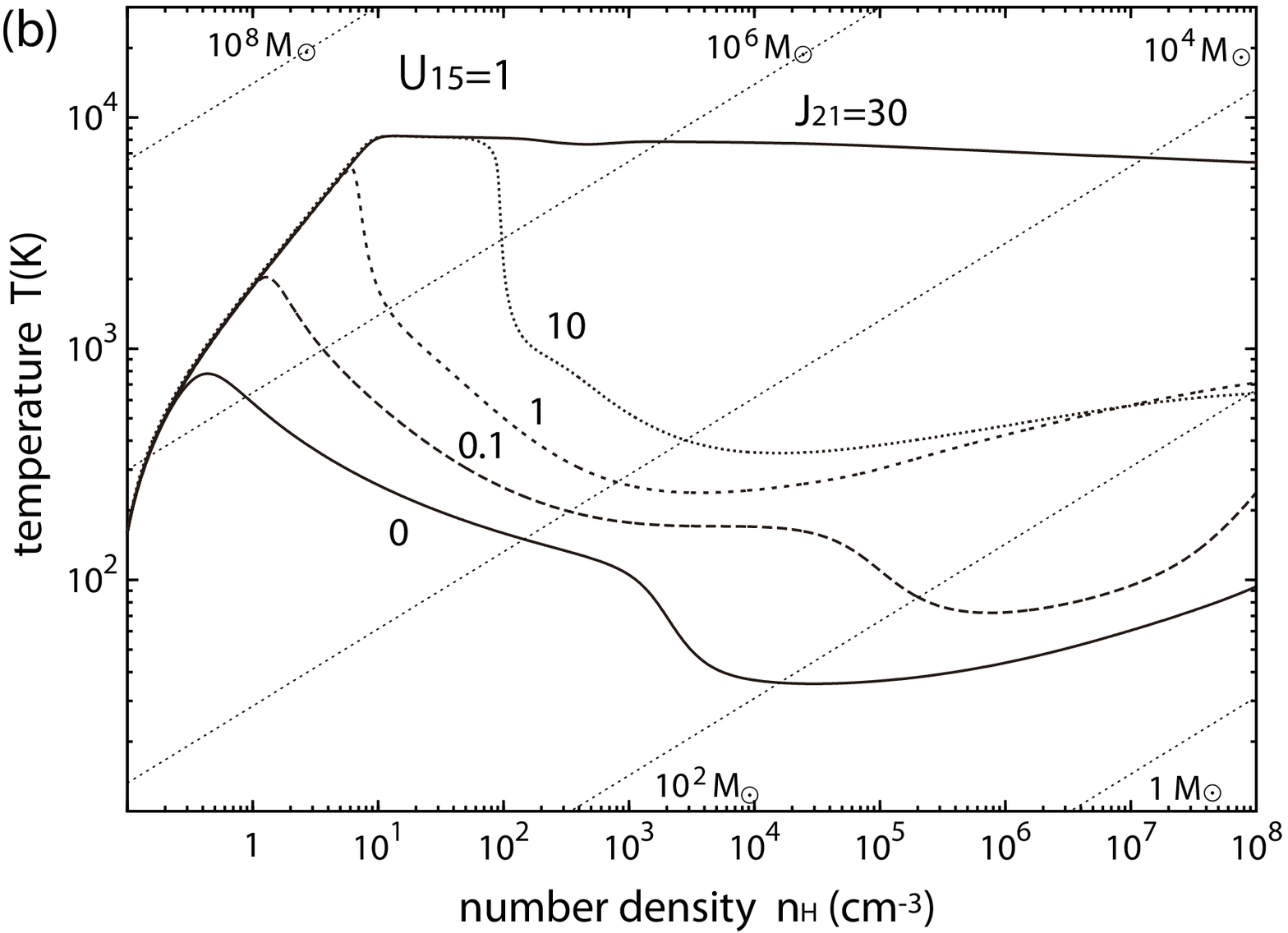}}
\\
\\
\rotatebox{0}{\includegraphics[height=59mm,width=80mm]{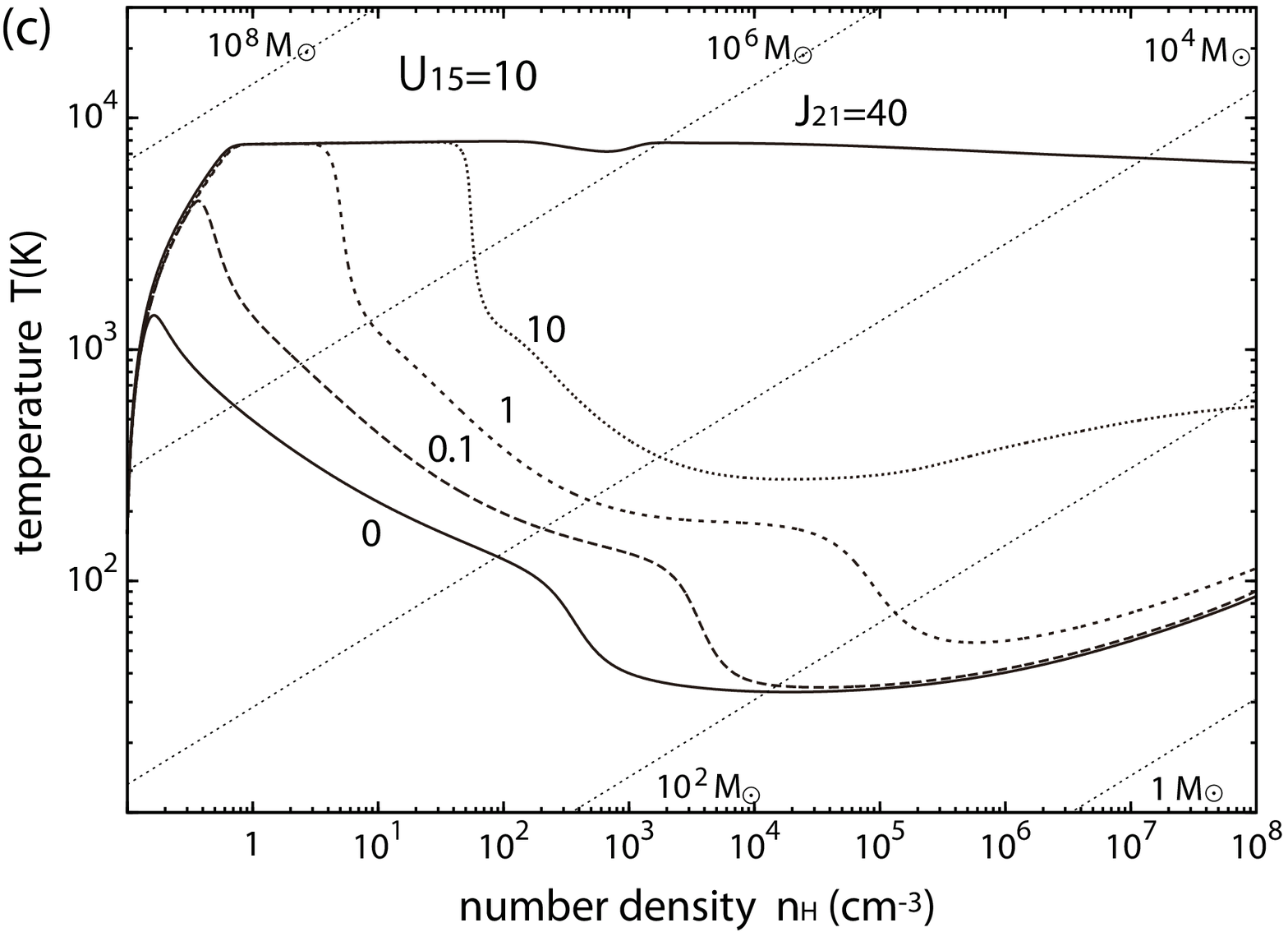}} \hspace{6mm}
\rotatebox{0}{\includegraphics[height=59mm,width=80mm]{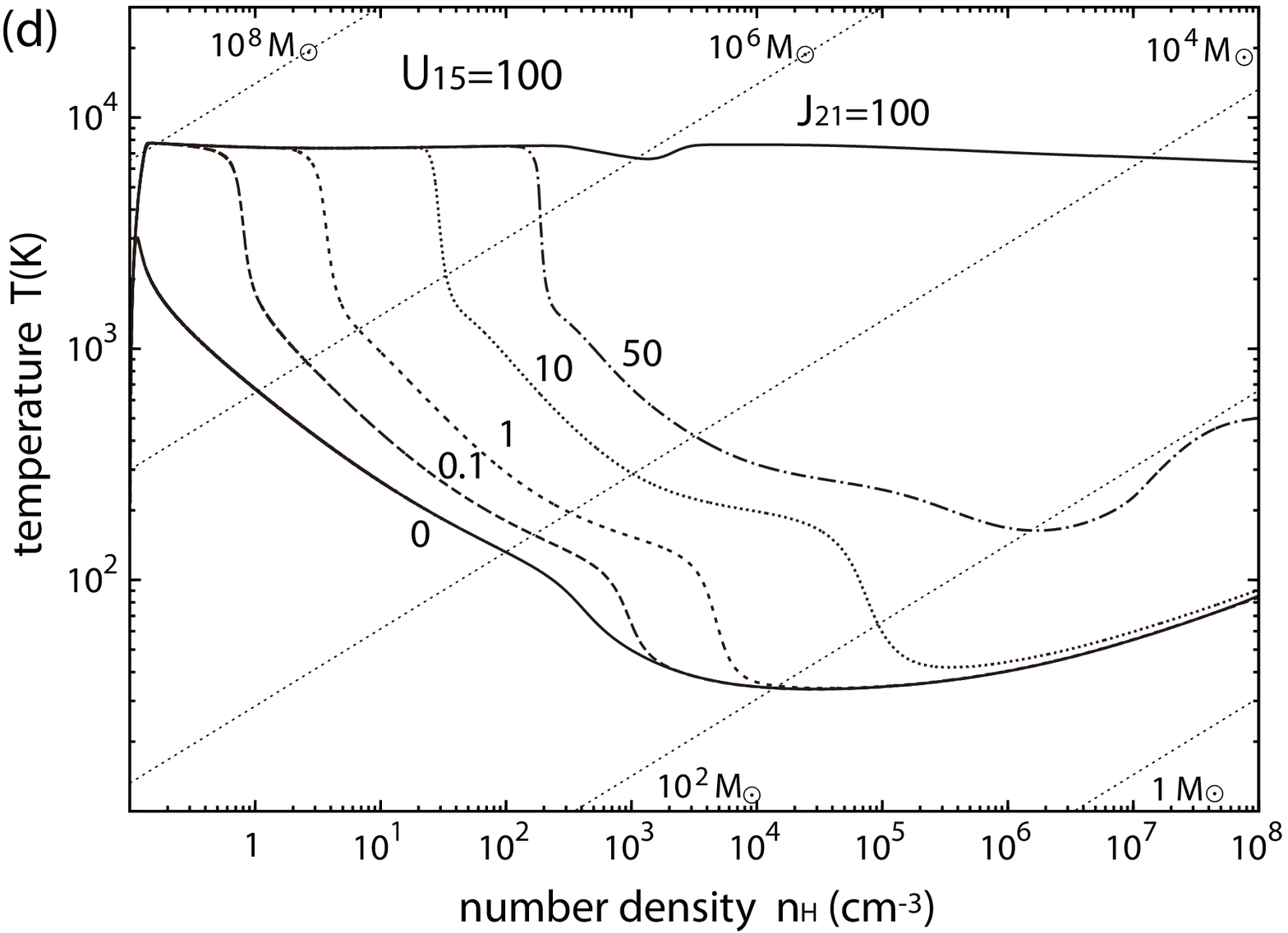}}
\\
\end{tabular}
}
\caption{Effect of cosmic rays on the temperature evolution of primordial-gas clouds under 
FUV irradiation with $T_{\ast}=10^{4}$ K.
Four panels correspond to the evolutionary tracks with four different 
cosmic-ray energy densities: $U_{15}=10^{-3}$, $1$, $10$ and $100$, where 
$U_{15}=U_{\rm{CR}}/10^{-15}$ erg cm$^{-3}$. 
The curves in each panel are those for different FUV fluxes, 
whose intensities $J_{21}$ are indicated by numbers.
The diagonal dotted lines show those at the constant Jeans mass, 
whose values are indicated by numbers in the Figure.
}
\label{fig:n-T_u}
\end{center}
\end{figure*}

\subsubsection{X-rays}
Since the ionization cross sections of hydrogen and helium fall as $\sigma _{\rm{H}}(\nu )\propto \nu ^{-3}$ 
and $\sigma _{\rm{He}}(\nu )\propto \nu ^{-2}$, respectively, 
towards higher energy photons, the soft X-ray ($2-10$ keV) photons 
reach far longer distance from the sources than ionizing UV photons.
Many sources contribute to X-rays at a given point, i.e., 
an extragalactic X-ray background is built up (Haiman, Rees \& Loeb 1997).

Following Glover \& Brand (2003), we assume
the incident X-ray background having a power-law spectrum with index $-1.5$, 
\begin{equation}
J_{\rm X}(\nu )=J_{\rm X, 21} \times 10^{-21}
\Big( \frac{\nu }{\nu _0}\Big) ^{-1.5} 
~\rm{erg \hspace{1mm} s^{-1} \hspace{1mm} 
cm^{-2} \hspace{1mm} sr^{-1} \hspace{1mm} Hz^{-1}}, 
\end{equation}
where $h\nu _0=1$ keV. 
We consider the X-ray ionization of both H and He atoms. 
The heating rate by X-rays is 
\begin{equation}
\Gamma _{\rm X}=\Gamma _{\rm X, H}+\Gamma _{\rm X, He}, 
\end{equation}
where 
\begin{equation}
\Gamma _{{\rm X}, i}=\int \frac{4\pi J_{\rm X}(\nu )}{h\nu }
e^{-\tau _\nu }
\sigma _{i}(\nu )
E_{{\rm h}, i} d\nu ~~(i={\rm H},~{\rm He}),
\end{equation}
the optical depth is given by 
\begin{equation}
\tau _\nu =N_{\rm H}\sigma _{\rm{H}}(\nu )+N_{\rm He}\sigma _{\rm He}(\nu ),
\end{equation}
$E_{{\rm h}, i}$ is the energy deposited as heat for each ionization process, 
and given by a formula by Wolfire et al. (1995).
The ionization rate is also expressed as
\begin{equation}
\zeta _{\rm X, p}^{i}=\int \frac{4\pi J_{\rm X}(\nu )}{h\nu }e^{-\tau _\nu }
\sigma _{i}(\nu )d\nu , 
\end{equation}
where the subscript p represents primary ionization by X-rays. 
The primary electron's energy is deposited not only in heating but also 
in the secondary ionization. 
Since the energy of the primary electron is far larger than the ionization 
threshold, 
the secondary ionization is more effective than the primary in the case of 
X-ray ionization. 
The secondary ionization rate of H is given by
\begin{equation}
\zeta _{\rm X, s}^{\rm H}=\Big( \zeta _{\rm X, p}^{\rm H}+\frac{y({\rm He})}{y({\rm H})}\zeta _{\rm X, p}^{\rm He}\Big) \langle \phi ^{\rm H}\rangle ,
\end{equation}
and the secondary ionization rate of He is expressed similarly,
where $\langle \phi ^{\rm H}\rangle $ is the number of secondary ionization 
per primary electron averaged over the X-ray spectrum, for which 
we use the expression by Wolfire et al. (1995). 
Then, the terms on the right hand side mean the secondary ionization rate 
by primary electrons due to H and He ionization, respectively. 
The total ionization rate by X-rays is given by the sum of the primary and 
secondary rates: 
\begin{equation}
\zeta _{\rm X}^{i}=\zeta _{\rm{X, p}}^{i}+\zeta _{\rm{X, s}}^{i} ~~(i={\rm H},~{\rm He}). 
\end{equation}
In considering the secondary ionization, the H and He ionization rates are about the same magnitudes, 
$\zeta _{\rm X}^{\rm H}\simeq \zeta _{\rm X}^{\rm He}$. 
We add the ionization rate by X-rays $\zeta _{\rm X}^{\rm H}$ ($\zeta _{\rm X}^{\rm He}$) to the photoionization rate $k_{21}$ ($k_{52}$, respectively).

\begin{figure}
\rotatebox{0}{\includegraphics[height=59mm,width=80mm]{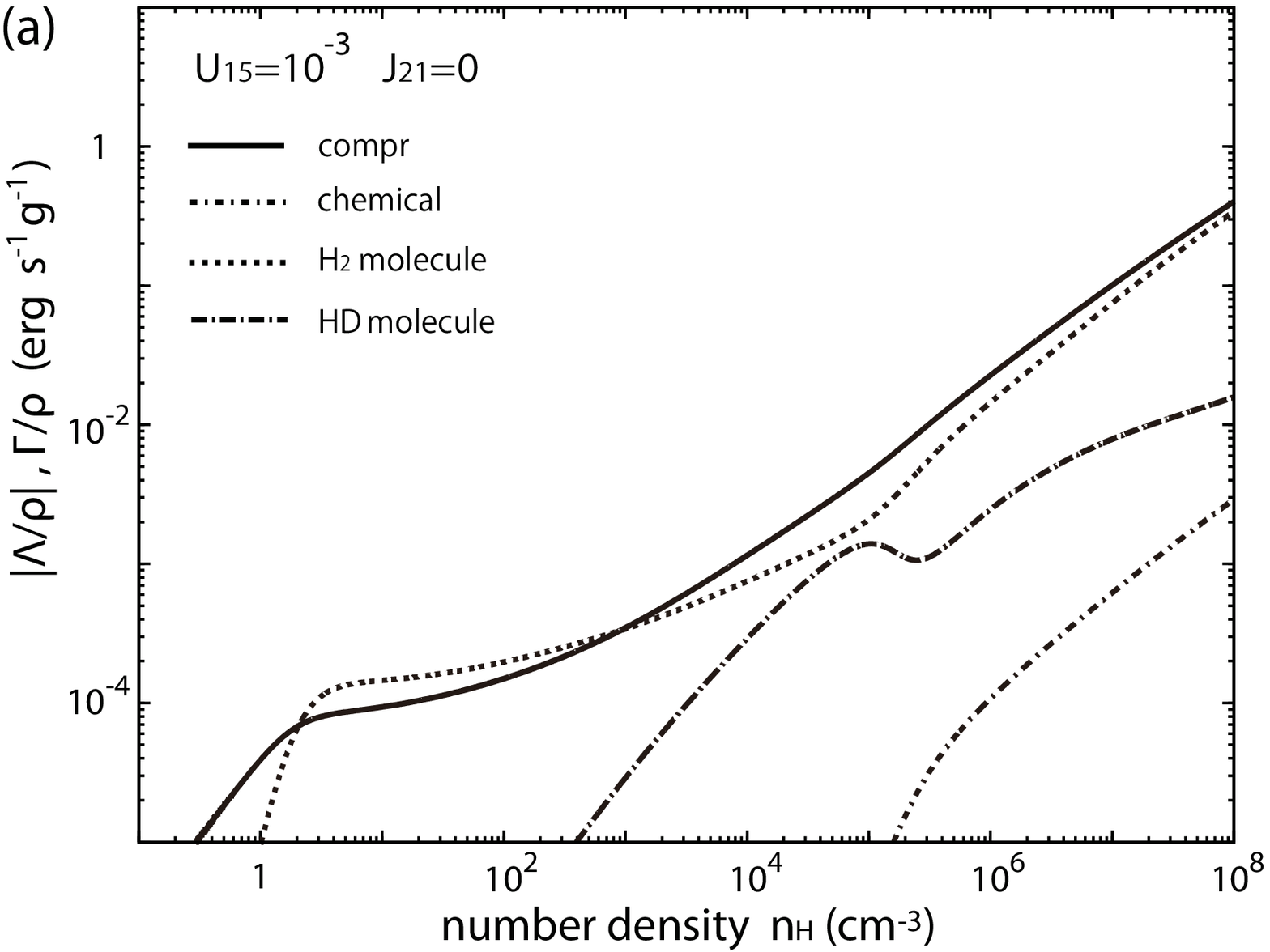}}
\\
\\
\\
\rotatebox{0}{\includegraphics[height=59mm,width=80mm]{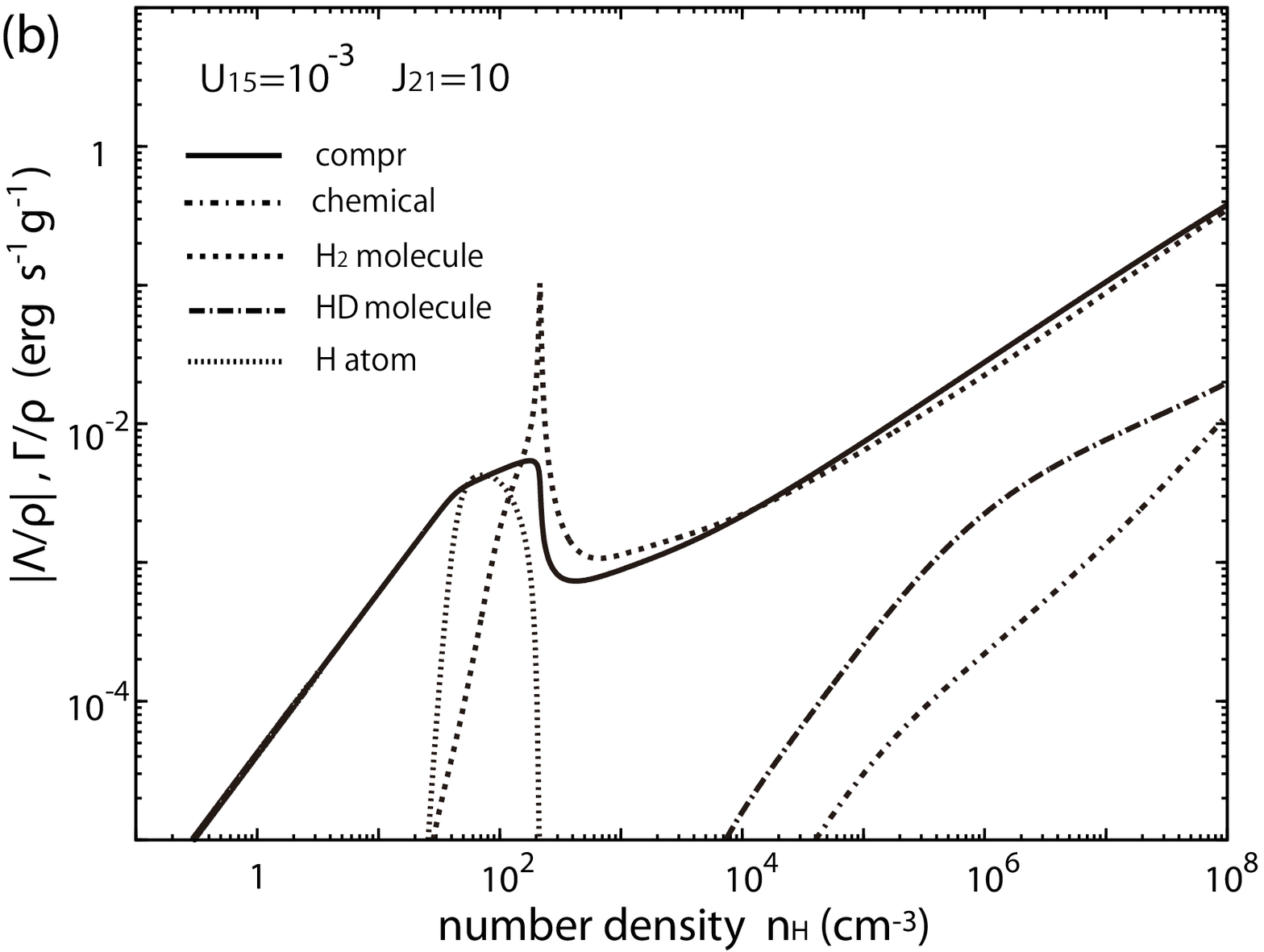}}
\\
\\
\\
\rotatebox{0}{\includegraphics[height=59mm,width=80mm]{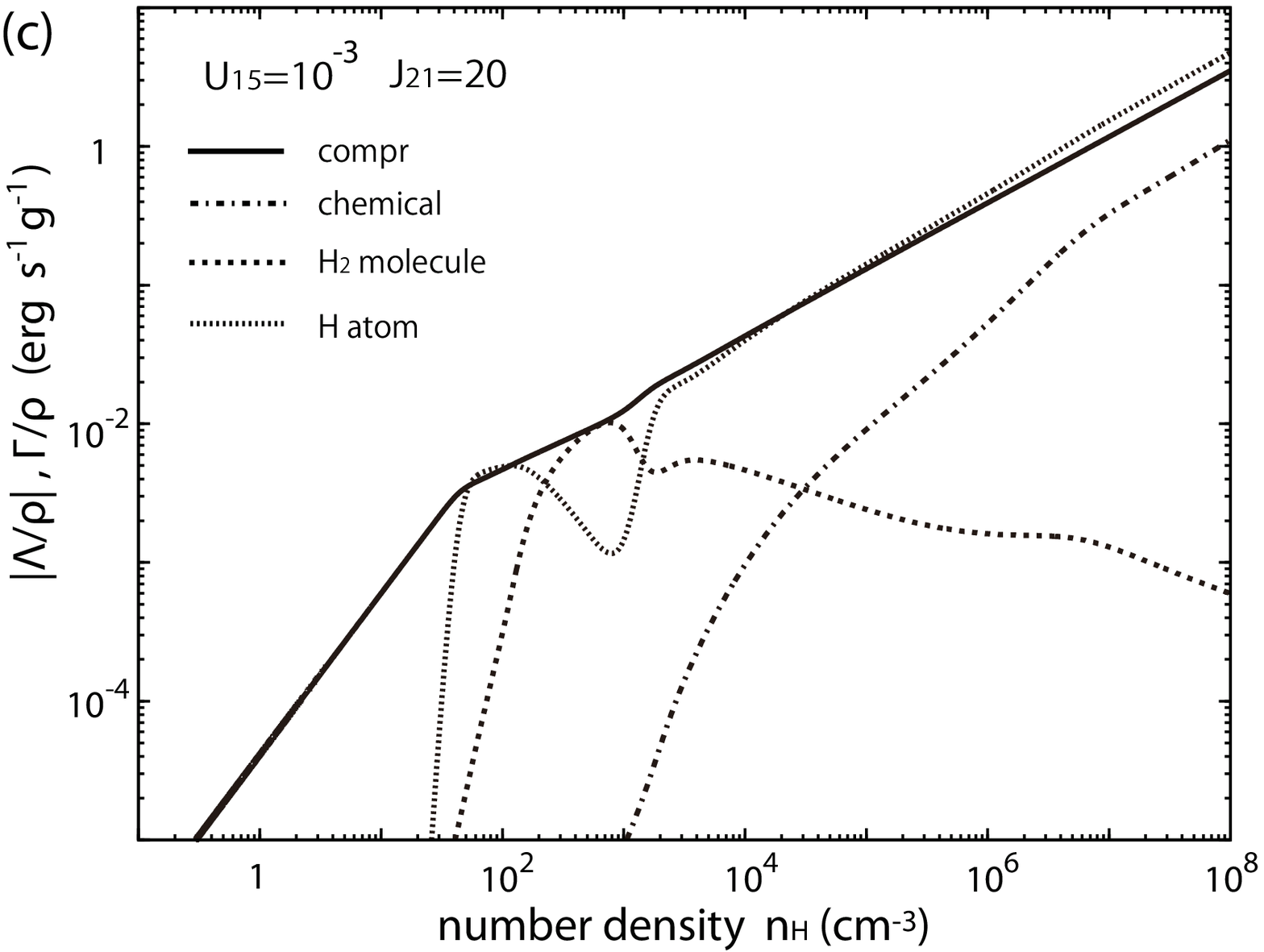}}
\caption{
Cooling and heating rates per unit mass as a function of the 
central number density. 
The cosmic ray energy density is $U_{15}=10^{-3}$, and the FUV intensity is 
(a) $J_{21}=0$, (b) $J_{21}=10$ and (c) $J_{21}=20$ for $T_{\ast}=10^4$ K.
The lines show the heating rates by compression (thick) and chemical reactions 
(short dash-dotted), and cooling rates by 
H$_2$ (thick dotted) and HD (dash-dotted) molecules, H atoms (dotted). 
}
\label{fig:c-h_u18}
\end{figure}
\begin{figure}
\rotatebox{0}{\includegraphics[height=59mm,width=80mm]{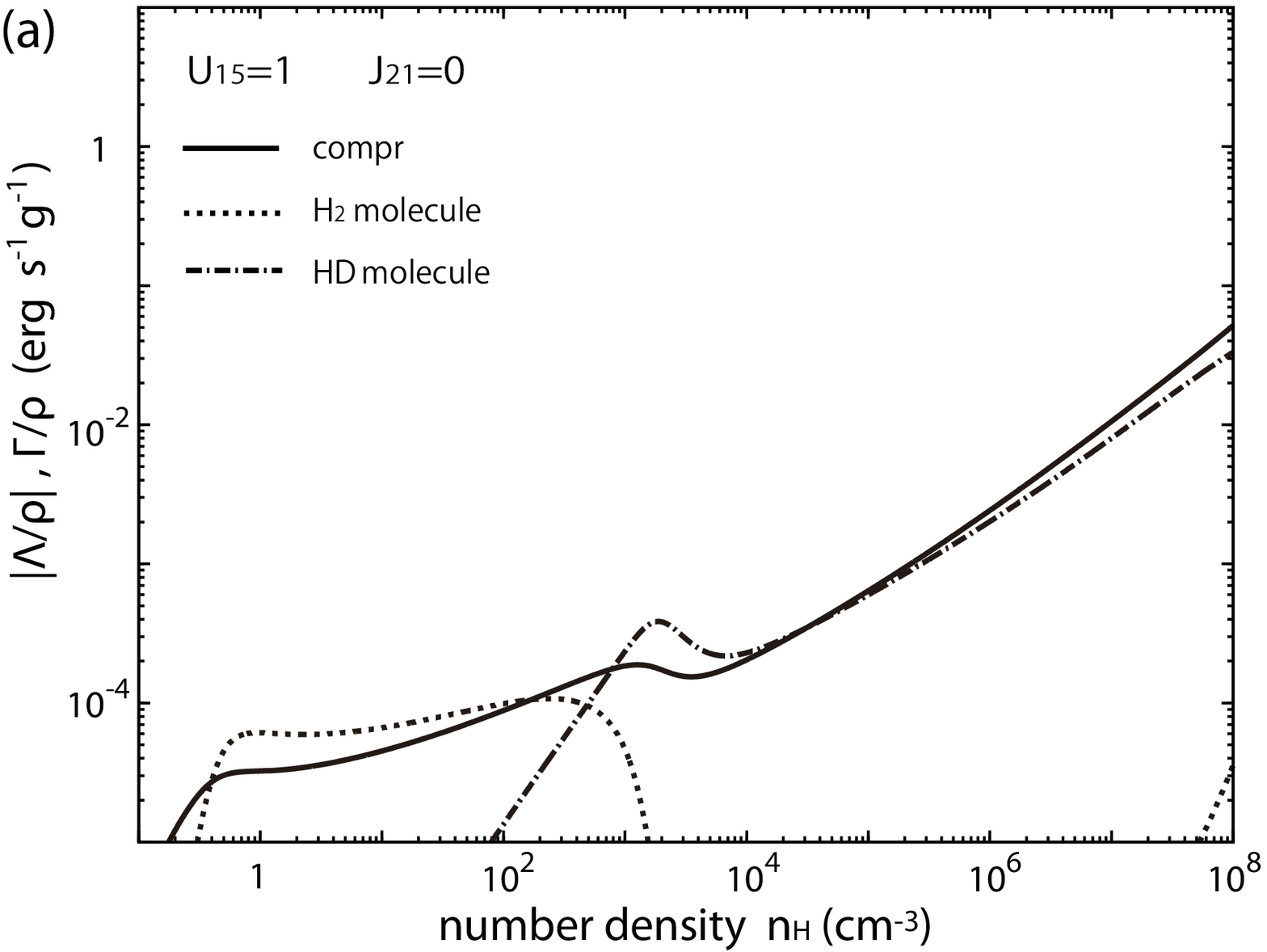}}
\\
\\
\\
\rotatebox{0}{\includegraphics[height=59mm,width=80mm]{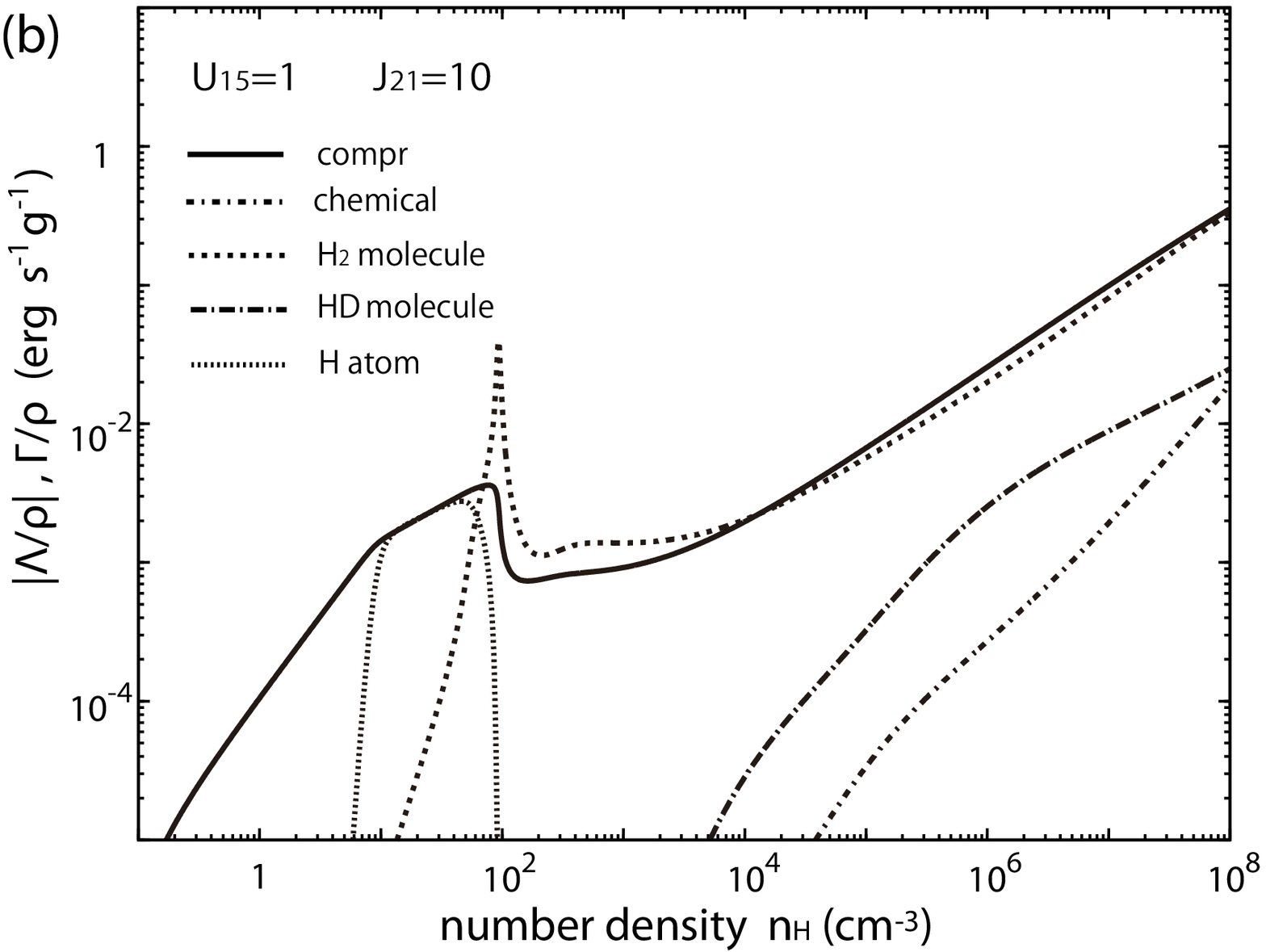}}
\\
\\
\\
\rotatebox{0}{\includegraphics[height=59mm,width=80mm]{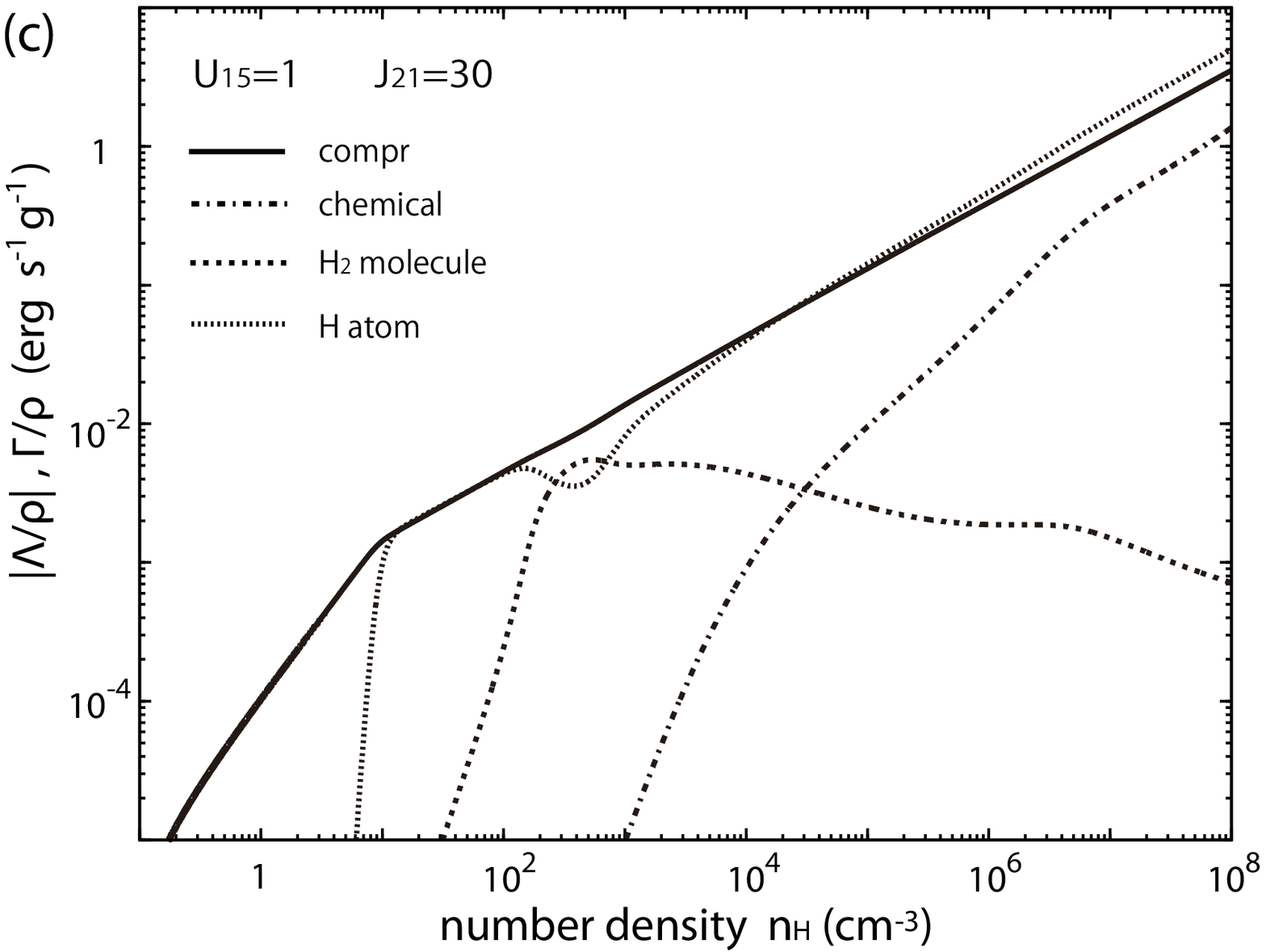}}
\caption{
The same as Fig.~\ref{fig:c-h_u18}, but for the cosmic ray energy density $U_{15}=1$, 
and the FUV intensity (a) $J_{21}=0$, (b) $J_{21}=10$ and (c) $J_{21}=30$.
}
\label{fig:c-h_u15}
\end{figure}
\begin{figure}
\rotatebox{0}{\includegraphics[height=59mm,width=80mm]{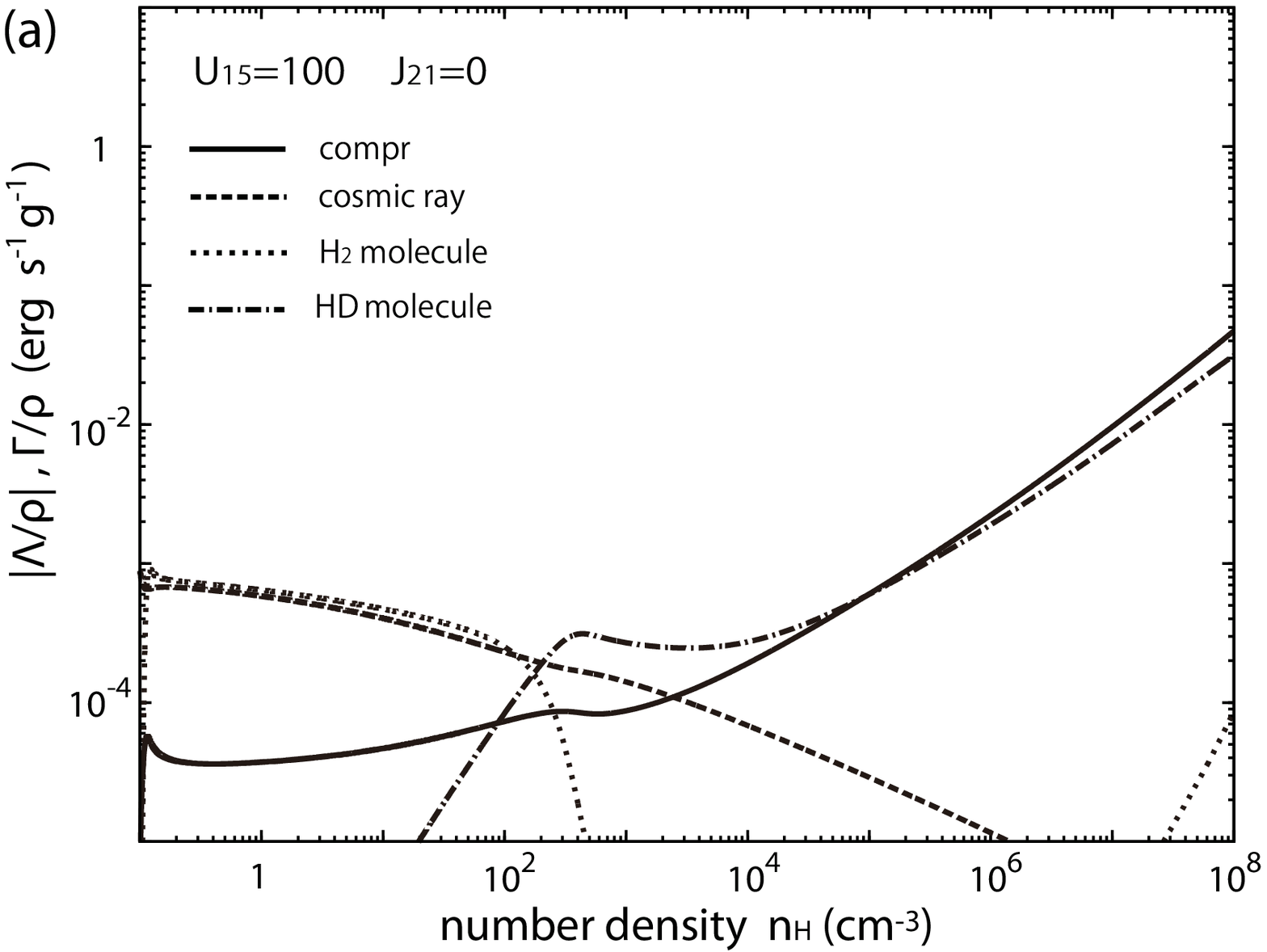}}
\\
\\
\\
\rotatebox{0}{\includegraphics[height=59mm,width=80mm]{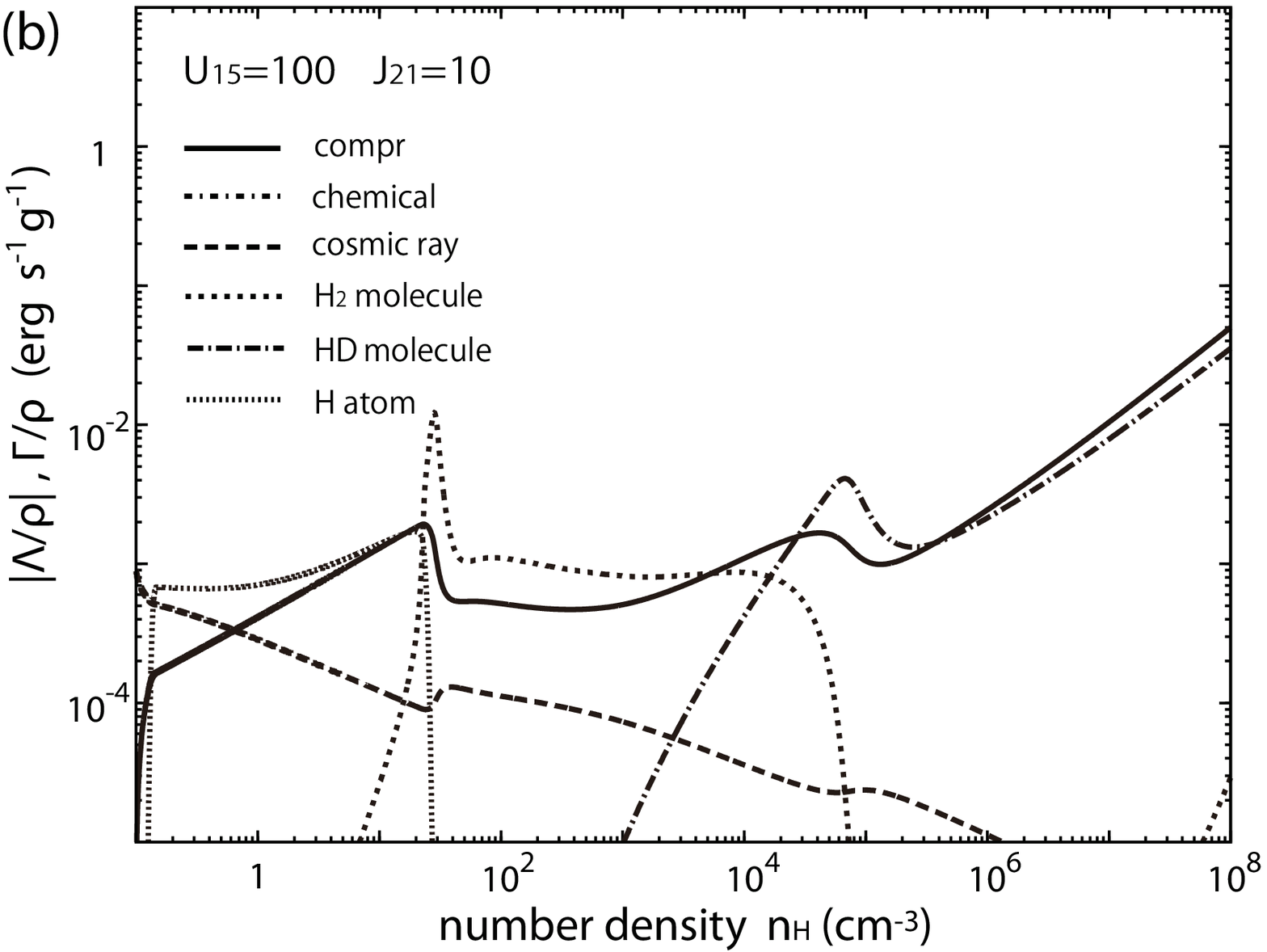}}
\\
\\
\\
\rotatebox{0}{\includegraphics[height=59mm,width=80mm]{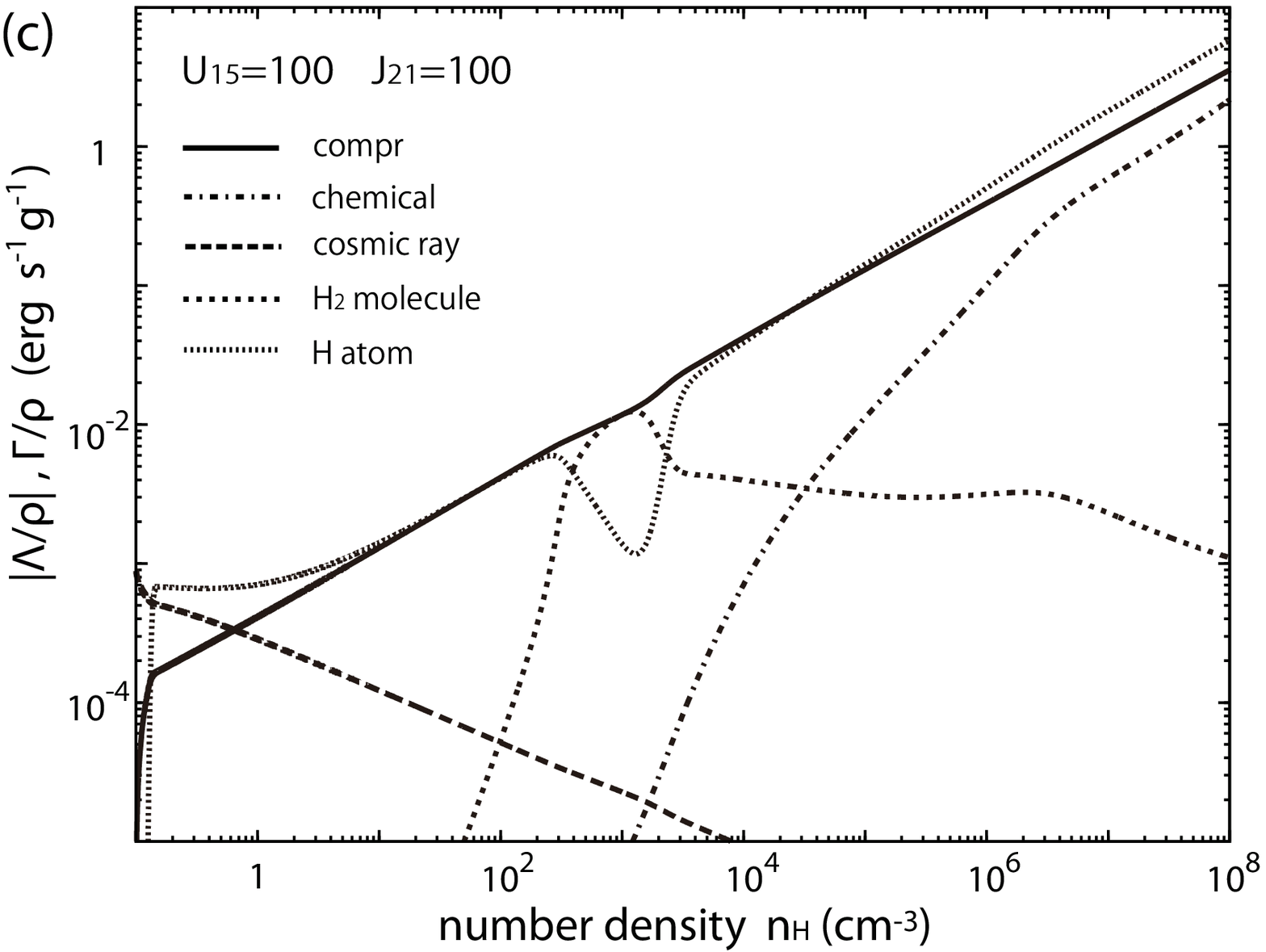}}
\caption{The same as Fig.~\ref{fig:c-h_u18}, but for the cosmic ray energy density $U_{15}=100$, 
and the FUV intensity (a) $J_{21}=0$, (b) $J_{21}=10$ and (c) $J_{21}=100$.
In addition to cooling and heating rates shown in Fig.~\ref{fig:c-h_u15}, 
heating rate by cosmic rays is also shown (dashed). 
}
\label{fig:c-h_u13}
\end{figure}

\section{Results}
We here present the results of our calculation and 
describe the physical processes determining the thermal evolution of the clouds.

\subsection{Effects of cosmic rays}
In Fig.~\ref{fig:n-T_u}, we show the temperature evolution of collapsing clouds 
irradiated by FUV fields with a diluted blackbody spectrum 
$T_{\ast}=10^4$ K, along with CRs with energy density 
(a) $U_{15}=10^{-3}$, (b) 1, (c) 10, and (d) 100. 
The lines in each panel are for cases with different values 
of FUV strength $J_{21}$, which are indicated by numbers in the panel. 
For the cases with $U_{15}=10^{-3}$, $1$, and $100$, we also plot 
in Figs~\ref{fig:c-h_u18}, \ref{fig:c-h_u15}, and \ref{fig:c-h_u13}, respectively, 
the cooling and heating rates per unit mass by individual processes, i.e., 
heating rates by compression (thick), by CR (dashed) and 
by chemical reactions (short dash-dotted), and 
cooling rates by the H$_2$ (thick dotted), 
by HD (dash dotted) and by H atom (dotted). 
Since the major cooling and heating processes for $U_{15}=10$ are 
very similar to those for $U_{15}=1$, we do not repeat presenting them. 

First we see the case of $U_{15}=10^{-3}$, where 
the CR is too weak to affect any aspect in the evolution 
(Figs~\ref{fig:n-T_u} a and \ref{fig:c-h_u18}). 
In the no-FUV case ($J_{21}=0$), the cloud collapses 
along the standard H$_2$-cooling track 
(e.g., Palla et al. 1983). 
Following the initial adiabatic rise up to $\sim 1000$ K, the temperature 
starts decreasing owing to the H$_2$ cooling, which is produced through 
the H$^{-}$ channel (reactions 3 and 4), until the critical density 
$\simeq 10^4$ cm$^{-3}$, where the H$_2$ level-populations reach the local 
thermodynamic equilibrium (LTE) and the cooling rate per unit mass saturates. 
The temperature thereafter increases again gradually by the compressional 
heating towards higher density (Fig.~\ref{fig:c-h_u18} a).
Addition of a FUV field affects the evolution in the following way. 
As seen in the cases with $J_{21}=0.1$ and 1, 
with increasing radiation intensity,  
the initial adiabatic phase continues 
until higher density and temperature, where 
the H$_2$ column density becomes high enough for efficient self-shielding 
against photodissociation.
Once the H$_2$ cooling becomes effective, the temperature decreases
and gradually converges to the track in the no-FUV case.
For $J_{21}=10$, before the onset of H$_2$ cooling,
the temperature reaches $\simeq 8000$ K and stays 
almost constant around $10^{2}$ cm$^{-3}$ owing to the H Ly $\alpha$ cooling 
(dotted line in Fig.~\ref{fig:n-T_u} a).
This short isothermal phase is followed by a rapid temperature drop
by the H$_2$ cooling (see also Fig.~\ref{fig:c-h_u18} b).
As in the cases with lower FUV fields,
the temperature converges towards the no-FUV track.
When a FUV field exceeds a critical value 
$J_{21}\geq J_{\rm{crit}}\simeq 20$, the temperature evolution 
is qualitatively altered (top solid line in Fig.~\ref{fig:n-T_u} a; 
see also Fig.~\ref{fig:c-h_u18} c).
The isothermal evolution at $\simeq 8000$ K continues until very high density 
($\sim 10^{16}~{\rm cm^{-3}}$) and the H$_2$ cooling never 
becomes important. 
This bifurcation of thermal evolution originates from the fact that if 
the H$_2$ formation is prevented until the critical density for LTE,
sufficient H$_2$ never forms because, at higher density, 
(i) the amount of H$_2$ needed for cooling increases 
(ii) collisional dissociation from the excited ro-vibrational levels of H$_2$ 
becomes effective, which reduces the H$_2$ fraction (Omukai 2001). 

\begin{figure*}
\begin{center}
\centerline{
\begin{tabular}{l l}
\rotatebox{0}{\includegraphics[height=59mm,width=80mm]{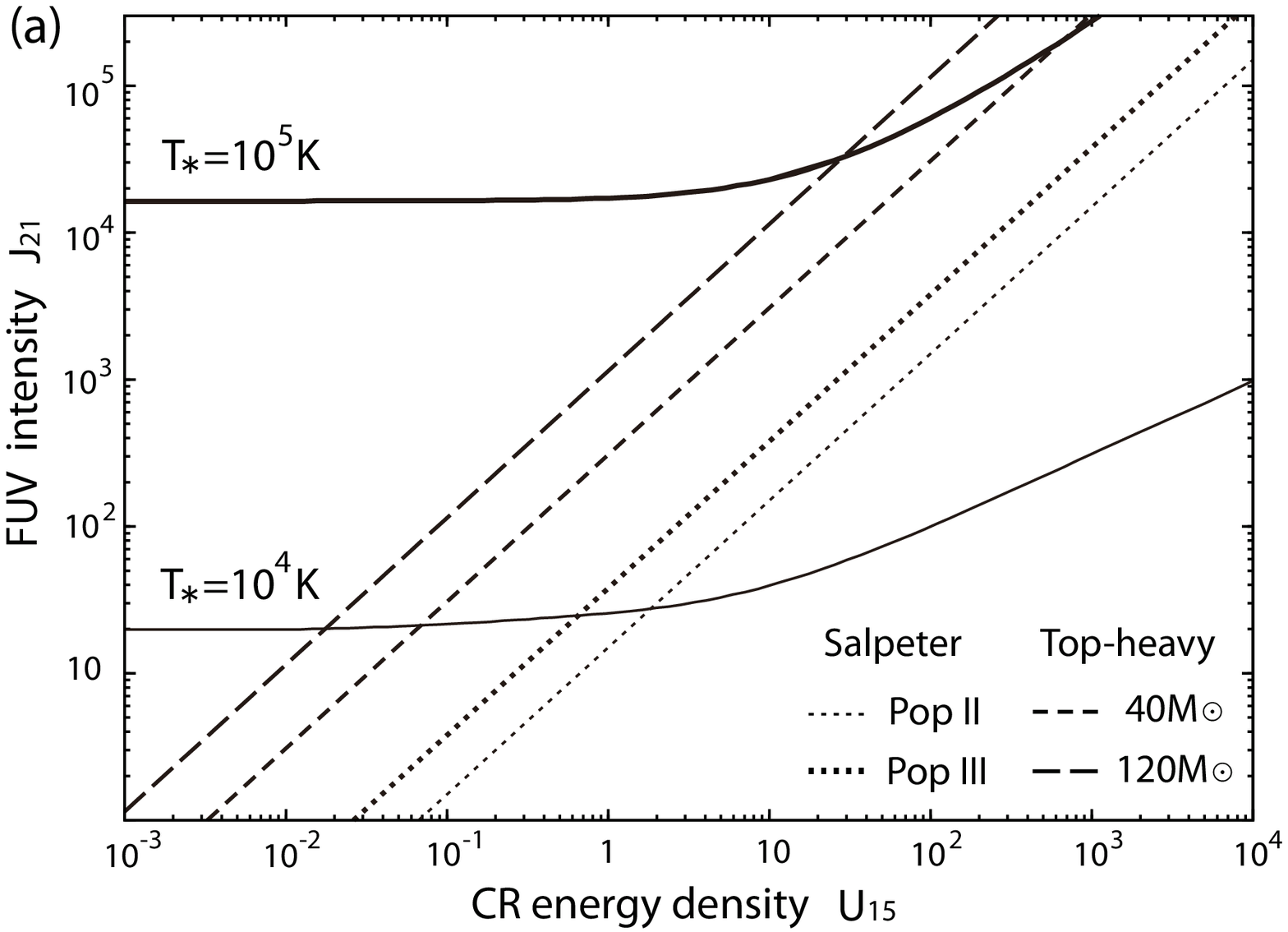}} \hspace{6mm}
\rotatebox{0}{\includegraphics[height=59mm,width=80mm]{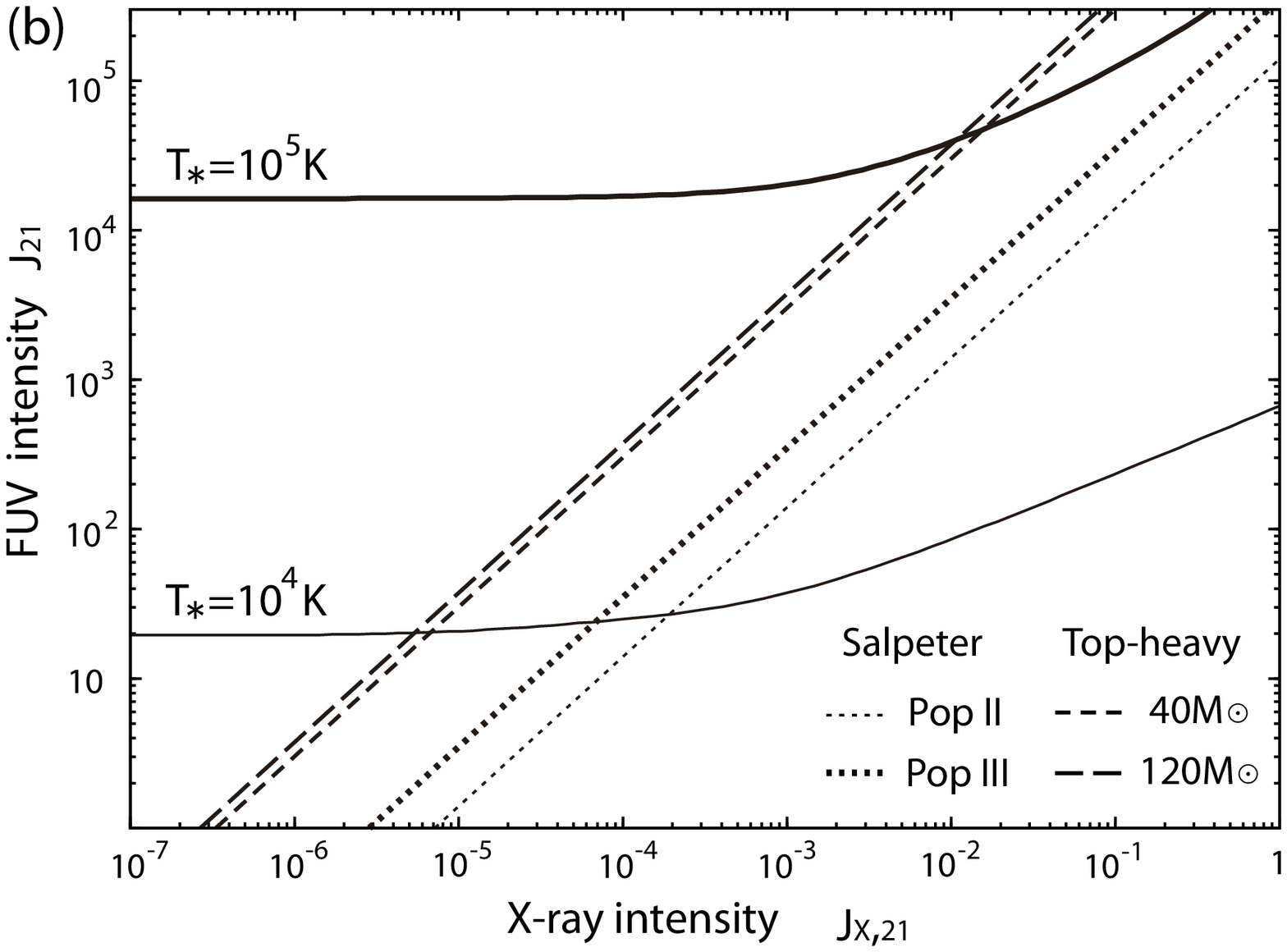}}
\end{tabular}
}
\caption{(a) The dependence of the critical FUV flux $J_{\rm crit}$ on the CR energy 
density $U_{15}$ (solid curves) for two spectral types (thin curve: $T_{\ast}=10^4$ K, 
thick curve: $T_{\ast}=10^5$ K). 
In both cases, $J_{\rm crit}\propto {U_{15}}^{1/2}$ for $U_{15}\ga 10$. 
The lines indicate the relations between $J_{21}$ and $U_{15}$ 
emitted from a star-forming galaxy (equation \ref{eq:ju}) 
with (i) the Salpeter IMF and $Z=10^{-3}~Z_{\sun }$ (Pop II; thin dotted), 
(ii) Salpeter IMF and $Z=0$ (Pop III; thick dotted), 
(iii) the top-heavy IMF with $m_{\rm OB}=40~{\rm M}_{\sun}$ (short dashed) 
and (iv) the top-heavy IMF with $120~{\rm M}_{\sun}$ (long dashed), respectively.
The intersection of the curves and lines 
for each $T_{\ast}$ case corresponds to the actual CR energy density and 
critical FUV flux $J_{\rm crit}$ in a halo affected by the star-forming galaxy. 
(b) the same plot for the X-ray ionization. 
For both values of $T_{\ast}$, $J_{\rm crit} \propto (J_{\rm X, 21})^{1/2}$ 
for $J_{\rm X, 21}\ga 10^{-3}$.
Note that the fraction of halos with $\ga 10^8~{\rm M}_{\sun}$ at 
$z=10$ irradiated by FUV radiation $J_{21}$ exceeding $\ga 10^5$ 
is negligibly low $\la 10^{-28}$.
}
\label{fig:ujxj}
\end{center}
\end{figure*}

Next, we see how the thermal evolution described above 
changes with increasing CR flux.
The CR effects are twofold, i.e., heating and ionization.
The stronger the CR flux is, the more rapid is the initial temperature-increase 
owing to the CR heating as seen in Fig.~\ref{fig:n-T_u} (b)$-$(d).
In Fig.~\ref{fig:c-h_u13} for $U_{15}=100$,  
very high CR-heating rate can be seen at the lowest density. 
Enhanced ionization degree facilitates the H$_2$ formation via H$^-$ channel.
To quench H$_2$ cooling totally, higher FUV flux is necessary, namely,       
the value of the critical FUV flux $J_{\rm crit}$ becomes larger.
For example, as seen in Fig.~\ref{fig:n-T_u} (d), under a very strong CR field 
of $U_{15}=100$, even such strong FUV flux as $J_{21}=50$ is not enough 
to quench the H$_2$ cooling. 
In Fig.~\ref{fig:ujxj} (a), we plot the critical FUV flux $J_{\rm crit}$ 
as a function of the CR intensity. 
In the case of $T_{\ast}=10^{4}$ K, from its low-CR limit of $J_{\rm crit}\simeq 20$, 
$J_{\rm crit}$ begins to increase around $U_{15}\sim 10$ and 
continues increasing as $J_{\rm crit} \propto {U_{15}}^{1/2}$ in the high $U_{15}$ limit.
The dominant cooling and heating processes are similar in all cases of 
the FUV flux exceeding the critical value $J_{\rm crit}$ 
(see Figs~\ref{fig:c-h_u18} c, \ref{fig:c-h_u15} c and \ref{fig:c-h_u13} c).

For low FUV flux $J_{21} \la 0.1-10$, the higher H$_2$ concentration by CR ionization results in 
the lower temperature than in the no-CR case (see Fig.~\ref{fig:n-T_u} b$-$d).
This low temperature environment ($\la 150$ K) favors HD formation, 
and its cooling causes further temperature decrease to a few 10 K 
(e.g., Fig.~\ref{fig:n-T_u} b$-$d for $J_{21}=0$). 
As seen in Fig.~\ref{fig:c-h_u15} (a), HD becomes the main cooling agent at high density and low temperature 
($n_{\rm H} \ga 10^3~{\rm cm^{-3}}$ and $T\la 150$ K).
Without extra ionization, enough HD for cooling is not produced 
under FUV radiation with $J_{21}\ga 10^{-2}$ for $T_*=10^4$ K
(Yoshida, Omukai \& Hernquist 2007; Wolcott-Green \& Haiman 2011). 
As seen in Figs~\ref{fig:n-T_u} (b) and \ref{fig:c-h_u15} (b), (c), 
for the case of $U_{15}=1$, the FUV flux higher than $J_{21}\geq 1$ suppresses 
the HD cooling 
and the thermal evolution becomes the same as that for $U_{15}=10^{-3}$. 
However, the presence of stronger CRs permits HD cooling despite of 
such a high FUV flux as $J_{21}\ga 10$ (see Fig.~\ref{fig:c-h_u13} b).  

So far, we limit our analysis to the radiation spectra of $T_{\ast}=10^{4}$ K. 
We also studied the case of $T_{\ast}=10^{5}$ K.
As mentioned in Section 2.4.1, the H$^-$ photodissociation rate 
depends sensitively on the brightness temperature $T_{\ast}$. 
Then, the concentration of H$_2$ produced through the H$^-$ channel 
(reactions 3 and 4) changes with the value of $T_{\ast}$. 
This leads to the enhancement of the critical FUV flux $J_{\rm crit}$ 
for $T_{\ast}=10^5$ K compared to the case for $T_{\ast}=10^4$ K (see Section 4).
The critical flux $J_{\rm crit}$ for $T_{\ast}=10^5$ K is also plotted in 
Fig.~\ref{fig:ujxj} (a). 
We find that $J_{\rm crit} = 1.6\times 10^4$ for $T_{\ast}=10^5$ K, 
while $J_{\rm crit}=20$ for $T_{\ast}=10^4$ K in the no-CR cases.
For CR energy density higher than $U_{15} \sim 10$, the critical FUV flux 
$J_{\rm crit}$ increases as $\propto {U_{15}}^{1/2}$ in both $T_{\ast}=10^{4}$ 
and $10^{5}$ K cases. 

\subsection{Effects of X-rays}
\begin{figure*}
\begin{center}
\centerline{
\begin{tabular}{l l}
\rotatebox{0}{\includegraphics[height=60mm,width=80mm]{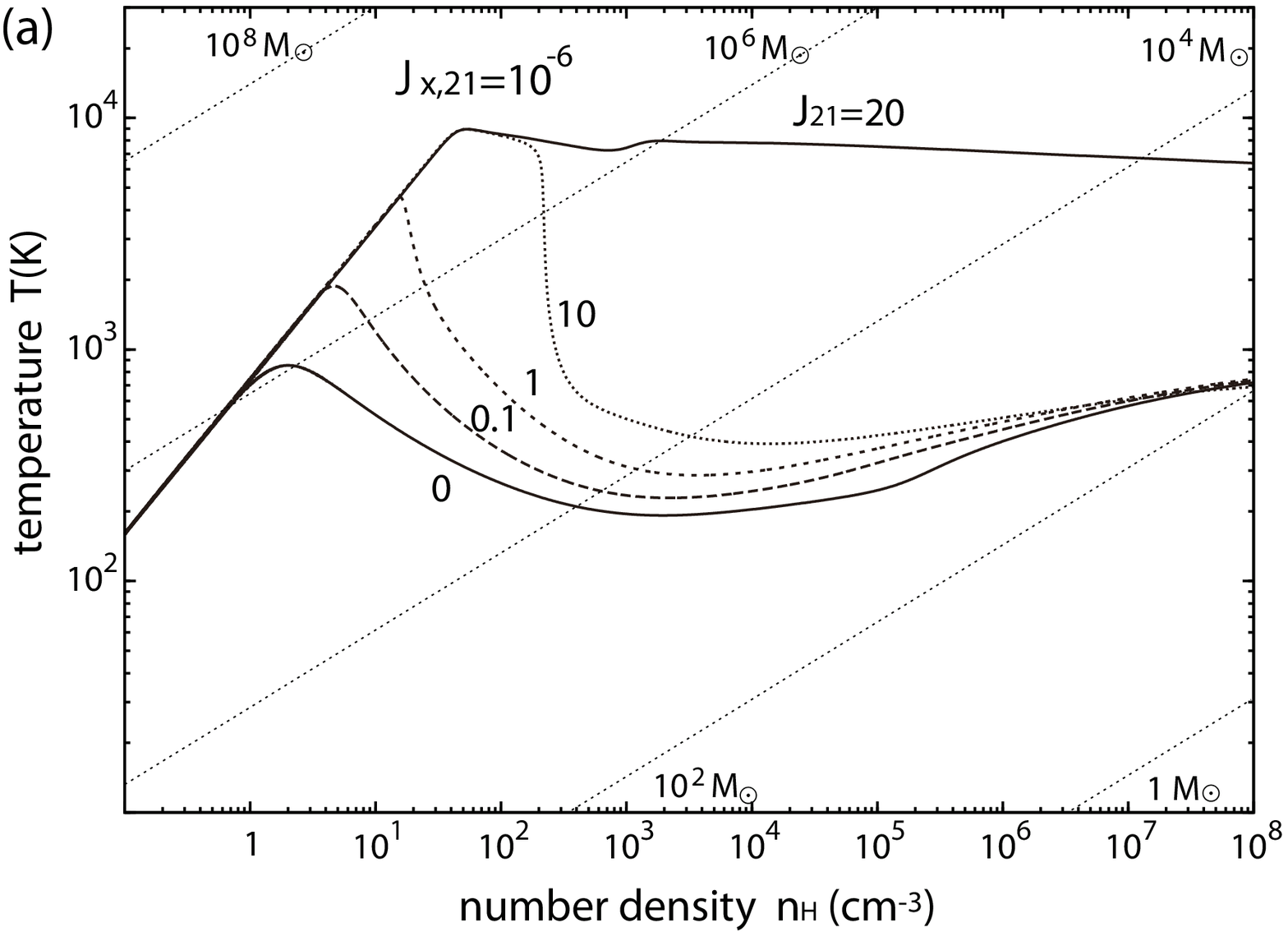}} \hspace{6mm}
\rotatebox{0}{\includegraphics[height=60mm,width=80mm]{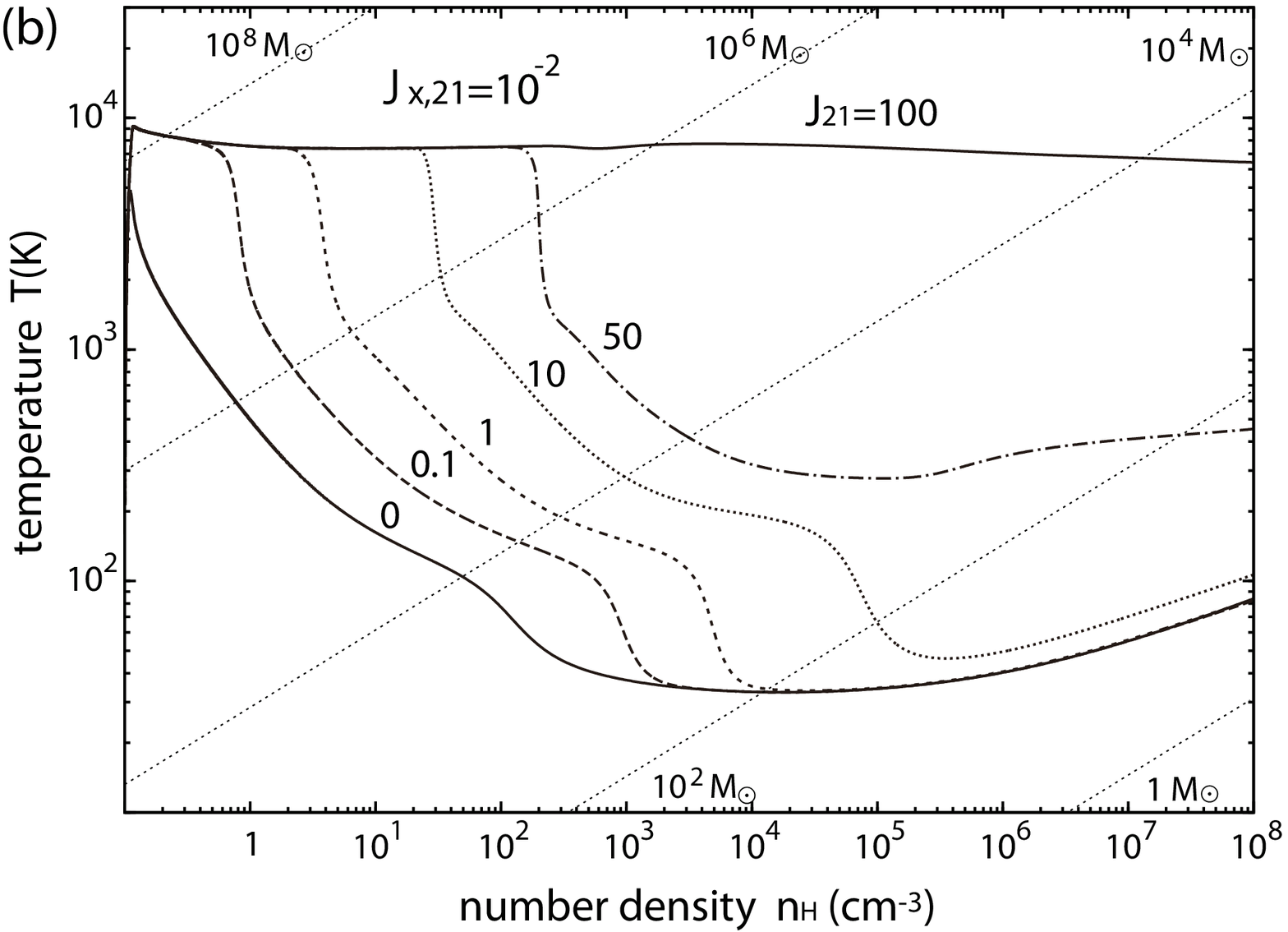}}
\end{tabular}
}
\caption{Effect of X-ray ionization on the temperature evolution of 
metal-free clouds under FUV irradiation with $T_{\ast}=10^4$ K.
Two panels show the evolutionary tracks with two different X-ray intensities 
$J_{\rm X, 21}=10^{-6}$ and $10^{-2}$. 
The curves in each panel are those for different FUV fluxes, whose intensities 
$J_{21}$ are indicated by numbers.
The diagonal dotted lines show those at the constant Jeans mass, 
whose values are indicated by numbers in the Figure.
}
\label{fig:n-T_x}
\end{center}
\end{figure*}

Here, we briefly mention the cases of the clouds irradiated 
both by FUV and X-rays. 
The temperature evolution is shown in Fig.~\ref{fig:n-T_x} 
for (a) $J_{\rm X, 21}=10^{-6}$ and (b) $10^{-2}$.
For the X-ray intensity as low as $J_{\rm X, 21}=10^{-6}$, 
the effects of X-rays, either 
ionization or heating, are not important and the evolutionary tracks in 
Fig.~\ref{fig:n-T_x} (a) are the same as those with negligible 
CRs ($U_{15}=10^{-3}$) in Fig.~\ref{fig:n-T_u} (a). 
When the X-ray intensity is elevated to $J_{\rm X, 21}=10^{-2}$, 
the gas is heated instantaneously (see Fig.~\ref{fig:n-T_x} b). 
In the case of $J_{21}=0$, the H$_2$ cooling balances the X-ray heating at 
$\simeq 5000$ K and the temperature begins decreasing thereafter. 
For $J_{21} \geq 0.1$, the temperature reaches at $\simeq 8000$ K 
and remains almost constant for a while 
by the H atomic cooling until H$_2$ is self-shielded against the FUV field. 
Similar to the case of CR ionization, X-ray ionization enhances 
the critical FUV flux $J_{\rm crit}$ for quenching the H$_2$ cooling. 
This critical FUV flux $J_{\rm crit}$ is plotted in Fig.~\ref{fig:ujxj} 
(b) as a function of the X-ray intensity $J_{\rm X, 21}$.
As in the case of CR ionization, $J_{\rm crit}$ remains constant 
below a threshold around $J_{\rm X, 21} \sim 10^{-3}$ and 
increases as $(J_{\rm X, 21})^{1/2}$ for higher X-ray intensity.  
Since the behaviors of temperature evolution for different FUV fluxes 
as well as the dominant cooling/heating processes are very similar 
to the cases with CRs, we do not repeat their description here.

\section{Dependence of the critical FUV flux on CR/X-ray intensity}

In Section 3, we see that strong CR and X-ray fluxes 
lead to the enhancement of the critical FUV flux as 
$J_{\rm crit}\propto {U_{15}}^{1/2}$ and 
$\propto (J_{\rm X, 21})^{1/2}$, respectively 
(see Fig.~\ref{fig:ujxj}).
Also, $J_{\rm crit}$ decreases with lowering FUV temperature $T_{\ast}$.  
In this section, we discuss the reason for these dependences. 

\begin{figure}
\rotatebox{0}{\includegraphics[height=63mm,width=85mm]{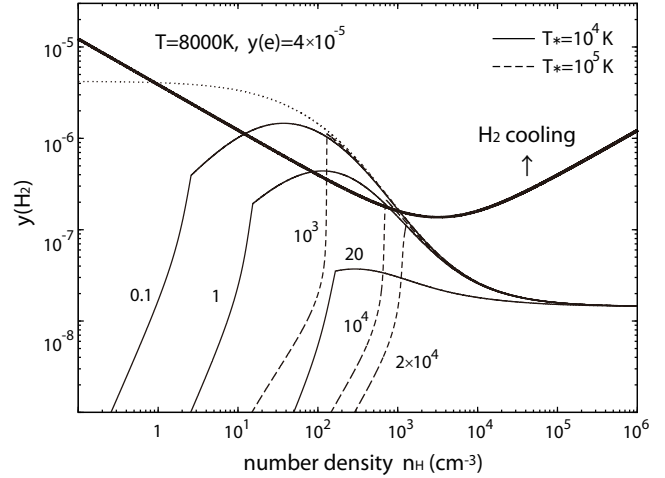}}
\caption{The H$_2$ concentration $y({\rm H_2})$ estimated by 
equation (\ref{eq:yH2}) (thin lines) vs. that needed for its cooling 
to exceed the compressional heating $y_{\rm cool}({\rm H_2})$ (thick line) 
in the isothermally collapsing clouds at $8000$ K with  
the ionization degree $y({\rm e^{-}})=4\times 10^{-5}$. 
The H$_2$ concentrations are shown for the FUV fields with brightness 
temperatures $T_{\ast}=10^4$ K (thin solid lines) and $10^5$ K (thin dashed lines).
For each value of $T_{\ast}$, three cases with different FUV intensities $J_{21}$, 
which are indicated by numbers in the Figure, are presented. 
The dotted line shows $y_{\rm cd}({\rm H}_2)$ without H$^-$ photodissociation.
}
\label{fig:H2}
\end{figure}

In the cloud under a high enough FUV field, the temperature 
reaches $\simeq 8000$ K and remains constant by the H Ly$\alpha$ cooling. 
If sufficient H$_2$ to cool forms during this isothermal collapse, 
the temperature drops by its cooling and the isothermal evolution 
is terminated. 
Otherwise, the isothermal evolution continues until very high density of  
$\sim 10^{16}~{\rm cm^{-3}}$, where the cloud becomes optically thick 
to the H$^{-}$ bound-free absorption (Omukai 2001). 
The H$_2$ concentration needed for its cooling to overcome
the compressional heating is
\begin{equation}
y_{\rm cool}({\rm H_2})=\frac{(3/2)k_{\rm B}T}{{\cal L}_{\rm{H}_2}t_{\rm{ff}}}, 
\end{equation}
where ${\cal L}_{\rm H_2} \equiv \Lambda_{\rm{H}_2}/n({\rm{H}}_2)$ 
(erg s$^{-1}$) is the cooling rate per an H$_2$ molecule.
Since most hydrogen is in the atomic form during the isothermal 
collapse, we assume $n({\rm H})=n_{\rm H}$ in this Section. 
In Fig.~\ref{fig:H2}, we show $y_{\rm cool}({\rm H}_2)$ (thick solid line) 
as a function of density for the isothermal collapse at $8000$ K.
Since ${\cal L}_{\rm H_2}$ increases linearly with number density 
until the critical density for LTE 
$n_{\rm H,cr} \simeq 3 \times 10^{3}~{\rm cm^{-3}}$ (at $8000$ K) and 
saturates for higher density, 
$y_{\rm cool}({\rm H_2})$ decreases as $\propto n_{\rm H}^{-1/2}$ 
until $n_{\rm H,cr}$, reaches  
the minimum $\simeq 1.4\times 10^{-7}$ 
there, and increases as $\propto n_{\rm H}^{1/2}$ for higher density. 
This $y_{\rm cool}({\rm H_2})$ is to be compared with 
the actual amount of H$_2$ present $y({\rm H_2})$. 
Before the temperature reaches $\simeq 8000$ K,
the H$_2$ concentration attains the equilibrium value set by 
the balance between the formation and destruction reactions. 
The effective rate coefficient of H$_2$ formation via H$^{-}$ channel 
taking account of the H$^{-}$ photodissociation is 
\begin{equation}
k_{\rm form}=k_{3}\frac{k_{4} n_{\rm H}}{k_{4} n_{\rm H}+k_{19}}, 
\end{equation}
where 
\begin{equation}
k_{19}=\beta_{19} J_{21}, 
\end{equation}
by using $\beta_{19} = 2 \times 10^{-7}~(1 \times 10^{-11})$ 
for $T_{\ast}=10^4$ K ($10^5$ K, respectively).
This H$_2$ is destructed either by photodissociation (reaction 18), which 
is dominant at the low density where the FUV radiation 
is not shielded, or collisionally (reaction 7) at higher density. 
Equating the larger of those dissociation rates with the formation rate,   
we obtain the H$_2$ concentration
\begin{align}
y({\rm H_2}) 
&={\rm min} (n_{\rm H}k_{18}^{-1}, k_{7}^{-1} ) 
k_{\rm form} y({\rm e}^-)
\nonumber \\
&\equiv {\rm min}(y_{\rm pd}({\rm H_2}), y_{\rm cd}({\rm H_2})). 
\label{eq:yH2}
\end{align}
The photodissociation rate coefficient is written as
\begin{equation}
k_{18}=\beta_{18} J_{21}, 
\end{equation}
where $\beta_{18}=4.2 \times 10^{-12}~(1.3 \times 10^{-12})$ 
for $T_{\ast}=10^4$ K ($10^5$ K, respectively).   
In the above, we assume the cloud is transparent 
to the photodissociating radiation
since $y_{\rm pd}({\rm{H}}_2)$ immediately exceeds 
$y_{\rm cd}({\rm{H}}_2)$ once the self-shielding becomes important. 
The values of $y({\rm{H}_2})$ by equation (\ref{eq:yH2}) are plotted 
in Fig.~\ref{fig:H2} for some combinations of $T_{\ast}$ and $J_{21}$ 
for the ionization degree $y({\rm e^-})=4 \times 10^{-5}$, 
corresponding the no-CR/X-ray case (see later).
The almost vertical portion of $y({\rm H_2})$ on the low-density side 
is limited by the photodissociation ($y_{\rm pd}({\rm H_2})$), 
while the gradually changing part 
at higher density is set by the collisional dissociation 
($y_{\rm cd}({\rm H_2})$). 
In the cases with $J_{21}=0.1$ and 1 ($10^{3}$ and $10^{4}$) 
for $T_{\ast}=10^{4}$ K ($10^{5}$ K, respectively), 
$y({\rm H_2})$ exceeds $y_{\rm cool}({\rm H_2})$ at some density.
At this moment, the actual temperature falls from $\simeq 8000$ K and 
the isothermal collapse terminates.  
On the other hand, in the case with $J_{21}=20$ ($2\times 10^{4}$) 
for $T_{\ast}=10^{4}$ K ($10^{5}$ K, respectively), $y({\rm H_2})$ always 
remains below $y_{\rm cool}({\rm H_2})$ and thus the isothermal collapse 
continues.

For density higher than the critical density 
$n_{\rm H, cr} \sim 10^{3}~{\rm cm}^{-3}$, 
the H$_2$ concentration decreases owing to effective collisional 
dissociation from excited ro-vibrational levels and has no chance 
to reach $y_{\rm cool}({\rm H_2})$ anymore.
To initiate efficient H$_2$ cooling, H$_2$ concentration must exceed 
$y_{\rm cool}({\rm H_2})$ before $n_{\rm H, cr}$. 
Thus, we can find the critical flux $J_{\rm crit}$ by the condition 
whether $y({\rm H_2})$ at $n_{\rm H, cr} \sim 10^{3}~{\rm cm}^{-3}$ is higher than 
$y_{\rm cool} ({\rm H_2})$. 

In the case of irradiation with hard FUV spectra (as $T_{\ast}=10^{5}$ K),
photodissociation limits the H$_2$ concentration and H$^{-}$ photodissociation is 
relatively irrelevant (i.e., $k_{\rm form} \simeq k_{3}$). 
From the condition $y_{\rm{pd}}({\rm{H}}_2)=y_{\rm cool}({\rm{H}}_2)$ 
at $n_{\rm H, cr}$,  
we obtain
\begin{equation}
J_{\rm crit}= \frac{1}{\beta_{18}}  
\frac{k_{3} n_{\rm H, cr} y({\rm e}^-)}{y_{\rm{cool}}({\rm{H}}_2)}.
\label{eq:jpd}
\end{equation}
On the other hand, for softer FUV spectra, (as $T_{\ast}=10^{4}$ K), 
the H$_2$ formation rate is significantly reduced by H$^{-}$ photodissociation: 
$k_{\rm form}=k_{3} k_{4} n_{\rm H}/k_{19}$. 
The dotted line in Fig.~\ref{fig:H2} shows $y_{\rm cd}({\rm H}_2)$ 
without H$^-$ photodissociation, which exceeds $y_{\rm cool}({\rm H}_2)$ before 
reaching $n_{\rm H, cr}$.
The actual value of $y_{\rm cd}({\rm H}_2)$ is, however, suppressed by H$^-$ photodissociation 
(see thin solid lines in Fig.~\ref{fig:H2})
and thus the H$_2$ concentration is limited by collisional dissociation. 
Therefore, we obtain from the condition $y_{\rm{cd}}({\rm{H}}_2)=y_{\rm cool}({\rm{H}}_2)$
at $n_{\rm H, cr}$:
\begin{equation}
J_{\rm crit}= \frac{k_{4}}{\beta_{19} k_{7}}  
\frac{k_{3} n_{\rm H, cr} y({\rm e}^-)}{y_{\rm{cool}}({\rm{H}}_2)}.
\label{eq:jcd}
\end{equation}
In general, the critical FUV flux is given by the larger of 
the above two values (equations \ref{eq:jpd} and \ref{eq:jcd}). 

Note that  $J_{\rm crit}$ increases with the ionization degree $y({\rm e}^-)$, 
which can be evaluated as follows. 
Without external ionization, 
the ionization degree is governed by 
\begin{equation}
\frac{d y({\rm e}^-)}{dt}= -k_{2} n_{\rm H} y({\rm e}^-)^{2}. 
\end{equation}
Integrating this equation for a collapsing cloud (i.e., equation \ref{eq:dndt}), 
\begin{equation}
y({\rm e}^-)=\frac{y_{0}({\rm e}^-)}{1+2k_{2} y_{0}({\rm e}^-) n_{\rm H} t_{\rm ff}},
\label{eq:ye0}
\end{equation}
where $y_{0}({\rm e}^-)$ is the initial ionization degree. 
For density $\ga 50~{\rm cm}^{-3}$, this renders 
$y({\rm e}^-)= (2k_{2}n_{\rm H}t_{\rm ff})^{-1}$.
Equation (\ref{eq:ye0}) gives $y({\rm e}^-) \simeq 4 \times 10^{-5}$ 
at $n_{\rm H}=10^{3}~{\rm cm}^{-3}$. 
Using the values at  $10^{3}~{\rm cm}^{-3}$, 
equations (\ref{eq:jpd}) and (\ref{eq:jcd}) 
lead to the critical FUV flux 
$J_{\rm crit} \sim 10$ ($10^4$) for $T_{\ast}=10^4$ K 
($10^5$ K, respectively), 
which are indeed similar to our numerical results for 
the no-CR/X-ray cases (Fig.~\ref{fig:ujxj}). 

In the presence of strong CRs, the ionization degree is set 
by the balance between recombination and CR ionization:
\begin{equation}
y({\rm e}^-)=\Big( \frac{\zeta _{\rm{CR}}}{k_2}\Big) ^{1/2}n_{\rm H}^{-1/2}.
\label{eq:ye1} 
\end{equation}
Comparing equations (\ref{eq:ye0}) and (\ref{eq:ye1}),
we see that the CR ionization controls the ionization degree if 
its rate is higher than 
\begin{equation}
\zeta _{\rm CR, crit}= \frac{1}{4k_2n_{\rm H}t_{\rm{ff}}^2}
\simeq 3.8\times 10^{-19}\hspace{2mm}{\rm{s}}^{-1}, 
\label{eq:ye2} 
\end{equation}
where the last equation is evaluated at $T=8000$ K.
This value is equivalent to the CR energy density 
$U_{\rm{CR, crit}}\simeq 5.5 \times 10^{-15}$ erg cm$^{-3}$ 
from equation (\ref{eq:zeta}).
For $U_{\rm CR} \ga U_{\rm CR, crit}$, 
the ionization degree, as well as the critical FUV flux, 
increases as $\propto {U_{\rm CR}}^{1/2}$.
This relation is reproduces well the behavior of $J_{\rm crit}$ 
as a function $U_{15}$ shown in Fig.~\ref{fig:ujxj} (a).

Note that our critical CR density is similar to another 
threshold value where the HD cooling becomes important 
in the absence of FUV radiation, 
$U_{\rm CR}=2 \times 10^{-15}~{\rm erg~cm^{-3}}$ 
(Stacy \& Bromm 2007). 
This is not a coincidence: 
both critical values are related to enhancement 
of the ionization degree and thus H$_2$ concentration 
from the case without CR ionization.  

Next we consider the case of the X-ray ionization.
Just the same as in the CR ionization, 
X-rays enhance the ionization degree 
as $y({\rm e}^-) \propto (J_{\rm X, 21})^{1/2}$ and thus 
$J_{\rm crit} \propto (J_{\rm X, 21})^{1/2}$
for $J_{\rm X, 21} \ga 10^{-3}$, whose threshold is 
determined by the balance between recombination and X-ray ionization.     
This describes well the results shown in 
Fig.~\ref{fig:ujxj} (b). 

Our value of $J_{\rm{crit}}=20$ for $T_{\ast}=10^4$ K is smaller than 
$J_{\rm{crit}}=39$ which Shang et al. (2010) estimated with one-zone model.
This difference comes from the H$_2$-formation rate 
at high temperature: 
the rate coefficient we adopt (Galli \& Palla 1998) 
is smaller than that Shang et al. (2010) used 
(Abel et al. 1997). 
On the other hand, our value of $J_{\rm{crit}}=1.6 \times 10^{4}$ 
for $T_{\ast}=10^5$ K is slightly larger than Shang et al. (2010) 
because we use the new shielding factor by Wolcott-Green \& Haiman (2011).

\section{CR and X-ray from star-forming galaxies}
So far, we have regarded the intensities of FUV, CRs and X-rays 
as free parameters in our calculation.
All those radiation fields are, however, closely linked to the 
star-formation activity in galaxies and thus their 
intensities, i.e., 
$J_{21}$, $U_{15}$ and $J_{\rm X, 21}$, are all
proportional to the star formation rate (${\rm SFR}$).
We here consider the star-forming galaxies with two types of initial mass 
function (IMF): the Salpeter IMF and the top-heavy one.
For the Salpeter IMF, we take the mass range of $1-100~{\rm M}_{\sun}$ and 
consider two cases of the stellar metallicity $Z=10^{-3}~Z_{\sun}$ and $0$, 
which corresponds to Pop II and III star clusters, respectively. 
For Pop III galaxies, we also consider the two cases of the top-heavy IMF 
where all stars are $m_{\rm OB}=40~{\rm M}_{\sun}$ or $120~{\rm M}_{\sun}$.
The IMF models we study are summarized in Table~2.

For FUV flux, we use the Lyman-Werner photon emissivity 
from a star-forming galaxy for the IMF models 
calculated by Schaerer (2002; 2003):
\begin{align}
J_{21}=\left\{ \begin{array}{ll}
6.5\times 10^2\\
1.6\times 10^3\\
1.4\times 10^4\\
1.7\times 10^4\\
\end{array} \right\}
\Big( \frac{d}{10~\rm{kpc}}\Big) ^{-2}
\Big( \frac{{\rm SFR}}{20~{\rm M}_{\sun}~\rm{yr}^{-1}}\Big),
\label{eq:jm0}
\end{align}
where $d$ is the distance from the source, and the numbers in the bracket
correspond to the IMF models in the same order as in Table 2. 

\begin{table}
\begin{center}
\caption{The IMF models of star-forming galaxies}
\begin{tabular}{c c c c c}
\hline
 model & IMF & mass (M$_{\sun}$) & metallicity ($Z_{\sun}$) & $T_{\ast}$ (K)\\ \hline \hline
(i) & Salpeter & $1-100$ & $10^{-3}$ & $10^4$\\
(ii) & Salpeter & $1-100$ & $0$ & $10^5$\\
(iii) & top-heavy & 40 & $0$ & $10^5$\\
(iv) & top-heavy & 120 & $0$ & $10^5$\\
\hline
\end{tabular}
\end{center}
\end{table}

We assume that supernova remnants (SNRs) are the sources of CRs. 
The total CR energy from a SNR is written as (Stacy \& Bromm 2007)
\begin{equation}
E_{\rm{CR}}= 10^{51} \Big( \frac{p_{\rm{CR}}}{0.1}\Big) 
\Big( \frac{E_{\rm{SN}}}{10^{52}~\rm{erg}}\Big) ~~{\rm erg}, 
\end{equation}
where $p_{\rm{CR}}$ is the fraction of SN explosion energy $E_{\rm{SN}}$ going 
into the CR energy.
Using the supernova rate in the source galaxy 
\begin{equation}
\dot{N}_{\rm SN}=\frac{f_{\rm OB}~{\rm SFR}}{\bar{m}_{\rm{OB}}}, 
\end{equation}
where $f_{\rm OB}$, $\bar{m}_{\rm{OB}}$ is the mass fraction and average mass of 
massive ($\geq 8~{\rm M}_{\sun}$) stars, 
the CR energy density of $U_{\rm{CR}}$ is
\begin{equation}
U_{\rm{CR}}=\dot{N}_{\rm{SN}}\frac{E_{\rm{CR}}}{4\pi d^2 \langle v\rangle }, 
\end{equation} 
where $\langle v\rangle$ is the average CR velocity.
For the fiducial values of $p_{\rm CR}=0.1$ and $E_{\rm SN}=10^{52}$ erg, we obtain
\begin{align}
U_{15}=\left\{ \begin{array}{ll}
42\\
42\\
44\\
15\\
\end{array} \right\}
\Big( \frac{d}{10~{\rm{kpc}}}\Big) ^{-2}
\Big( \frac{\rm SFR}{20~{\rm M}_{\sun}~{\rm{yr}}^{-1}}\Big).
\label{eq:um0}
\end{align}
From equations (\ref{eq:jm0}) and (\ref{eq:um0}), 
\begin{align}
J_{21}=\left\{ \begin{array}{ll}
~~~~15\\
~~~~38\\
3.1\times 10^2\\
1.4\times 10^3\\
\end{array} 
\right\}
U_{15}.
\label{eq:ju}
\end{align}

X-rays are mainly emitted by high-mass X-ray binaries. 
Observationally, the X-ray luminosities in $2-10$ keV of 
local star-forming galaxies are related with their
star formation rates as (e.g., Glover \& Brand 2003) 
\begin{equation}
L_{\rm X}=1.2\times 10^{40}
\Big( \frac{{\rm SFR} }{20~{\rm M}_{\sun}~{\rm{yr}}^{-1}}\Big) 
~~{\rm{erg}}\hspace{1mm}{\rm{s}}^{-1},
\end{equation}
namely,  
\begin{equation}
J_{\rm X, 21}=4.5\times 10^{-3}\Big( \frac{d}{10~\rm{kpc}}\Big) ^{-2}
\Big( \frac{{\rm SFR}}{20~{\rm M}_{\sun}~\rm{yr}^{-1}}\Big) .
\label{eq:xm0}
\end{equation}
Although this is an empirical relation for local galaxies, 
observations of Lyman-break galaxies provide 
supports for this to be valid as far as $z\sim 4$ 
(Glover \& Brand 2003 and references therein).
Here, we thus extend the relation (\ref{eq:xm0}) to high-redshift 
(say, $z \sim 10$) universe.
From equations (\ref{eq:jm0}) and (\ref{eq:xm0}), we obtain
\begin{align}
J_{21}=\left\{ \begin{array}{ll}
1.4\times 10^5\\
3.5\times 10^5\\
3.0\times 10^6\\
3.7\times 10^6\\
\end{array} \right\}
J_{\rm X, 21}.
\label{eq:jx}
\end{align}

In Fig.~\ref{fig:ujxj} (a), we plot the relations (\ref{eq:ju}) 
for the IMF cases (i)-(iv) in Table~2.
The FUV intensity and CR energy density from galaxies are 
expected to fall on this relation. 
In the case of Pop II radiation sources (i.e., type i), 
which are characterized by 
$T_{\ast} \sim 10^4$ K, the relation (\ref{eq:ju}) 
intersects with the $J_{\rm crit}$ curve at  
$(J_{21},~U_{15})=(20,~2)$, where the $J_{\rm crit}$ does not 
deviate yet from its no-CR value. 
This means that the CR ionization does not modify the value of $J_{\rm crit}$ significantly. 
On the other hand, for Pop III star clusters ($T_{\ast} \sim 10^5$ K), 
the critical FUV flux at the intersection points are $\ga 20$ (type ii), $10$ (iii), 
and $2$ times (iv) as large as $J_{\rm crit}$ in the no-CR case, respectively. 
Namely, the CR ionization drastically changes the critical FUV flux 
except for the case with the top-heavy IMF with $m_{\rm OB}=120~{\rm M}_{\sun}$.

In Fig.~\ref{fig:ujxj} (b), we also plot the relations (\ref{eq:jx}) 
between $J_{21}$ and $J_{\rm X, 21}$. 
As in the case of CRs, the X-ray ionization effects become important 
for higher $T_{\ast}$, in particular, $T_{\ast} \ga 10^4$ K. 

\section{Conclusion and Discussion}
We have calculated the thermal evolution of primordial clouds under 
strong far-ultraviolet (FUV) fields, along with cosmic rays (CRs)/X-rays.
In the cloud under a FUV field exceeding a threshold, 
the H$_2$ cooling is suppressed at any density.  
Such a cloud collapses almost isothermally at $8000$ K by hydrogen atomic cooling. 
According to numerical simulations (Bromm \& Loeb 2003), 
it avoids fragmentation and collapses directly to a supermassive star (SMS). 

Without external ionization by CRs/X-rays, the critical FUV flux is 
$J_{\rm crit} \simeq 1.6\times 10^{4}$ (in units of $10^{-21}~{\rm erg~s^{-1}~cm^{-2}~sr^{-1}~Hz^{-1}}$) 
for a diluted black body spectrum with 
brightness temperature $T_{\ast}=10^{5}$ K, 
while it is as small as $J_{\rm 21} \simeq 20$ for $T_{\ast}=10^{4}$ K. 
This dependence on $T_{\ast}$ comes from the higher H$^-$ photodissociation ($>0.755$ eV) rate for lower $T_{\ast}$ at the same 
$J_{\rm 21}$ normalized at $13.6$ eV. 
Since H$_2$ is produced through the H$^-$ channel (reactions 3
and 4 in Table 1), 
the larger H$^-$ photodissociation reduces the amount of H$_2$ and 
thus the value of $J_{\rm crit}$. 

We have studied how this critical FUV flux changes with 
simultaneous irradiation of either CRs or X-rays. 
In the case of CR irradiation, the critical FUV flux $J_{\rm crit}$ begins 
to increase with the CR energy density $U_{\rm CR} \ga 10^{-14}~{\rm erg~cm^{-3}}$
and depends as $J_{\rm crit} \propto {U_{\rm CR}}^{1/2}$ asymptotically. 
Similarly, for X-ray irradiation, $J_{\rm{crit}}\propto {J_{\rm X}}^{1/2}$ 
for $J_{\rm X} \ga 10^{-24}~{\rm erg~s^{-1}~cm^{-2}~sr^{-1}~Hz^{-1}}$ at $1$ keV. 
In both cases, the increase of the critical FUV flux is caused by 
the enhanced ionization by CRs/X-rays, which promotes the H$_2$ formation and cooling.   

FUV intensity and CRs/X-rays from star-forming galaxies are expected to 
correlate each other since they all trace the massive-star forming activity. 
Using the expected relations between FUV intensity and CRs/X-rays, 
we have found that if the initial mass function (IMF) of the radiation source is Salpeter-like and 
the brightness temperature of the FUV radiation is rather high $T_{\ast} \sim 10^5$ K, 
the critical FUV intensity increases significantly 
$J_{\rm crit} \sim 10^{6}$ owing to ionization by CRs and X-rays. 
Even with the top-heavy IMF, the critical FUV intensity increases to 
$J_{\rm crit}\sim 10^{5}$ unless the stellar mass is $\ga 100~{\rm M}_{\sun}$. 
Since the fraction of halos exposed to FUV flux exceeding $J_{\rm crit}$ 
decreases exponentially with $J_{\rm crit}$ for $\ga 2\times 10^3$ 
(Dijkstra et al. 2008),
there is little possibility ($\la 10^{-28}$) for such intense FUV field realized 
in any halos. 
We conclude that if the radiation source is composed of Pop III 
stars with brightness temperature $\sim 10^5$ K and contains 
sources of CRs/X-rays, its IMF must be very top-heavy $\ga 100~{\rm M}_{\sun}$
to enable SMS formation in nearby halos.
Other possible radiation sources enabling SMS formation include, e.g., 
sources with low brightness temperature, $T_{\ast}\simeq 10^4$ K, 
such as Pop II/I star clusters, or those with high $T_{\ast}$ 
but without strong CRs and X-rays. 
The latter possibility is realized if a Pop III star cluster 
is so young that it harbors neither supernova remnants nor high-mass X-ray binaries.

We have also found that the extragalactic 
ionization effects (CRs/X-rays) are able to promote the HD formation and cooling 
even if the clouds are irradiated with FUV radiation.
Without external ionization, Yoshida et al. (2007) and Wolcoott-Green \& Haiman (2011) 
showed that the HD formation is suppressed by FUV radiation 
with $J_{21}\ga 10^{-2}$ for $T_{\ast }=10^4$ K,
and that the number of metal-free stars ($\ga 10~{\rm M}_{\sun}$) formed 
in H{\scriptsize II} regions by HD cooling (so-called Population III.2 stars) is reduced. 
However, we find the possibility that such stars can form under
FUV radiation if strong CRs/X-rays ionize the clouds.
These mean that the star formation in H{\scriptsize II} regions by HD cooling is 
not suppressed so strongly as previous thought (see also Wolcott-Green \& Haiman 2011).

Finally, we discuss limitations of our analysis. 
We have treated the cloud evolution at the center
by the one-zone model, where two assumptions have been made.
First one is that the density increases 
approximately at the free-fall rate (equation \ref{eq:dndt}). 
This is a good approximation for the isothermal spherically symmetric collapse. 
In reality, however, in addition to the thermal pressure, 
the turbulent support and compression, as well as the centrifugal support, 
which we have not taken into account, can have importance in dynamics, 
which controls the collapse rate.
For example, in relatively massive halos, which we have considered 
here, the primordial gas is known to be in a state of turbulence 
(Wise \& Abel 2007; Greif et al. 2008). 
Such turbulence might affect the thermal evolution, as well as the 
exact value of the FUV critical intensity for suppressing H$_2$ cooling.
If the angular momentum is present,  
the centrifugal force becomes more important with contraction 
relative to the gravity, and eventually halts the collapse.
We here, however, only considered a rather low-density regime, 
and thus the angular momentum effect is not so dynamically important 
in our case. 
Another assumption in our model is that the external radiation is attenuated 
with the column density estimated by the central density and the size of the core, 
which is given by the Jeans length (equation \ref{eq:opt}).
Although the column density of the core is given by this value for 
the Larson-Penston solution, the contribution from the envelope needs to be added.
But we can see the envelope contribution is not significant by the following consideration. 
The steep decline of density with radius as $\rho \propto r^{-2}$ limits  
the column density of hydrogen nuclei $N_{\rm H}$ in the envelope at most that of the core. 
Due to the photodissociation and low formation rate, the H$_2$ abundance 
is remarkably lower in the envelope than in the core. 
Therefore, the envelope contribution to the column density of H$_2$, which is relevant for 
evaluating the critical FUV intensity, is small and 
the H$_2$ column density is well approximated by that of the core.
Another concern is about the geometry of the cloud. 
In the low-density regime we considered, 
the cloud shape can strongly deviate from the sphere.     
Even though the clouds are in such shapes as sheet-like or filamentary,
however, the length scale of the shortest axis, which is most relevant 
for the shielding effects, is still roughly given by the Jeans length, 
and thus the our assumption of the Jeans length shielding remains valid. 
In fact, Shang et al. (2010), who studied
the evolution of the primordial clouds 
under strong FUV irradiation and compared the results by 
the one-zone model and those by the cosmological three-dimensional 
simulation, found very good agreements as for 
the thermal evolution at the center of the clouds, as well as for the critical 
FUV intensity $J_{\rm crit}$ needed to quench the H$_2$ formation/cooling. 
In addition, we assume that the clouds are isotropically 
irradiated by FUV, as well as CRs and X-rays.
However, such intense FUV field as exceeding the critical value $J_{\rm crit}$ tends to 
be dominated by a nearby single large source, rather than by a collective effect 
of a large number of sources (Dijkstra et al. 2008).
We suspect that, in the case of such anisotropic radiation field,  
more intense sources are necessary to induce SMS formation as 
the radiation comes from only limited solid angles. 
To confirm it, however, detailed modeling is required, which is 
beyond the scope of this paper. 

\section*{Acknowledgments}
We would like to thank Takashi Nakamura for his continuous encouragement,  
and Takashi Hosokawa and Susumu Inoue for fruitful discussions. 
This work is supported by the Grant-in-Aid for the Global COE Program 
"The Next Generation of Physics, Spun from Universality and Emergence" 
from the Ministry of Education, Culture, Sports, Science and Technology (MEXT) of Japan. 
K.I. is supported by the Grants-in-Aid for the Japan Society for the Promotion of Science 
Fellows (23$\cdot$838). 
K.O. is supported by the Grants-in-Aid by the Ministry of Education, 
Science and Culture of Japan (19047004, 2168407, and 21244021).

\bsp

\label{lastpage}

\end{document}